\documentstyle[aps,prb,eqsecnum,multicol,epsbox,epsf]{revtex}
\def\mbf(#1){\mbox{\boldmath $#1$}}
\begin{document}
\draft
\preprint{HEP/123-qed}

\title{Electronic and Magnetic Properties of 
Nanographite Ribbons}

\author{Katsunori Wakabayashi}
\address{
Yukawa Institute for Theoretical Physics, Kyoto University, 
Kyoto 606-8502, Japan \\
and 
Institute of Materials Science, University of Tsukuba, 
Tsukuba 305-8573, Japan}
\author{Mitsutaka Fujita\footnote{deceased}}
\address{
Institute of Materials Science, University of Tsukuba, Tsukuba 305-8573,
 Japan}
\author{Hiroshi Ajiki}
\address{
Depertment of Material Physics, Osaka University, Osaka 560-8531, Japan}

\author{Manfred Sigrist}
\address{
Yukawa Institute for Theoretical Physics, Kyoto University, 
Kyoto 606-8502, Japan}

\date{\today}
\maketitle

\begin{abstract}
Electronic and magnetic properties of ribbon-shaped nanographite
systems with zigzag and armchair edges
in a magnetic field are investigated by using
a tight binding model.
One of the most remarkable features of these systems
is the appearance of edge states, 
strongly localized near zigzag edges.
The edge state in magnetic field, generating a rational fraction of the
magnetic flux ($\phi= p/q$) in each hexagonal plaquette of the
graphite plane, behaves like a zero-field edge state 
with q internal degrees of freedom.
The orbital diamagnetic susceptibility strongly depends on
the edge shapes. 
The reason is found in the analysis of the ring currents,
which are very sensitive to the
lattice topology near the edge. 
Moreover, 
the orbital diamagnetic susceptibility is scaled as a
function of the temperature, Fermi energy and ribbon width.
Because the edge states lead to 
a sharp peak in the density of states at the Fermi level,
the graphite ribbons with zigzag edges show
Curie-like temperature dependence of the Pauli paramagnetic 
susceptibility. 
Hence, it is shown that
the crossover from high-temperature diamagnetic
to low-temperature paramagnetic  behavior 
of the magnetic susceptibility 
of nanographite ribbons with zigzag edges.
\end{abstract}

\pacs{73.20At, 81.50.Tp, 75.20.-g}

\begin{multicols}{2}[]

\section{Introduction}
Nanographites are nanometer-sized graphite fragments which represent
a new class of a mesoscopic systems intermediate between aromatic
molecules and extended graphite sheets. In these systems the boundary
regions play an important role so that edge effects may influence
strongly the $\pi$-electron states near the Fermi energy. 
A useful and simple system to investigate the electronic states of
nanographites is provided by ribbon-shaped graphite sheets. 
The study of the electronic states of graphite ribbons based on the
tight binding model reveals that the edge shape - we distinguish
between {\it zigzag} and {\it armchair} edges (see Fig.1) - leads to a 
strikingly different properties of the states
near the Fermi level.  The ribbons with zigzag edges possess
partly flat bands at the Fermi level corresponding to electronic states 
localized in the near vicinity of the edge. These localized
states (``edge states'')  correspond to the non-bonding molecular
orbital (NBMO) as can be seen examining the analytic solution for
semi-infinite graphite with a zigzag edge\cite{peculiar,nakada}. 
In contrast localized edge states and the corresponding flat bands are 
completely absent for ribbons with armchair edges.

The localized edge states are of special interest in nanographite
physics, because of 
their relatively large contribution to the density of
states (DOS) at the Fermi energy. There is a tendency towards a Fermi
surface instability which is important to examine. Previously, it was
reported that based on the Su-Schrieffer-Hegger (SSH) model the
electron-phonon  interaction does not induce a lattice distortion
because of the non-bonding character of the edge states\cite{SSH}. 
On
the other hand, the electron-electron interaction on the level of an
unrestricted Hartree-Fock approximation (HFA) yields a ferrimagnetic
spin polarization at the zigzag edges and an energy gap at the Fermi
level\cite{peculiar,RPA}. The effect of long-range Coulomb interaction
on the edge state was  examined using the Par-Parier-Pople (PPP) model
with the restricted HFA which does not allow any  spin
polarization. The conclusion was that long-range Coulomb interaction
does not destroy the edge state as well  as the flat bands\cite{PPP}. 
This results was further confirmed by first principle calculations
based on local density approximation (LDA)\cite{miyamoto}.

The stability of the edge states has been extensively
investigated from various points of view. One remaining problem is 
the influence of an external magnetic field. 
How are the edge states affected?
Is their NBMO character preserved?
The answers to these questions will be important for future studies of 
the electronic, magnetic and transport properties of nanographites.
For this purpose we investigate here the magnetic properties,
especially magnetic susceptibility, for the case of non-interacting
electrons.

The observed magnetic susceptibility $\chi$
is the sum of four components: (1) localized spin 
susceptibility $\chi_{spin}$, (2) diamagnetic susceptibility due to
the core  electrons $\chi_{core}$, (3) Pauli paramagnetic
susceptibility $\chi_{P}$  and (4) orbital diamagnetic
susceptibility $\chi_{orb.}$ due to the cyclotron motion of the
itinerant electrons.

Since we neglect electron-electron interaction throughout this paper,
$\chi_{spin}$ can be neglected. Furthermore, $\chi_{core}$ is
unimportant for us, because it is small and basically temperature
independent. On the other hand, the Pauli paramagnetic susceptibility
is related to the DOS 
at  the Fermi level, which represents an important component
in zigzag nanographite ribbons where an enhanced density of states
appears at the Fermi level. Note that $\chi_{P}$ is negligible
in armchair ribbons, aromatic molecules and graphite sheets, because
their DOS is suppressed at the Fermi level.
We will see below that the fact that the DOS introduced by the edge
states is sharply peaked at the Fermi energy, $\chi_{P}$ introduces
a very pronounced temperature dependence which is nearly Curie-like.
The diamagnetic contribution to the susceptibility is very familiar
from the magnetic properties of graphite sheets. It is due to the
orbital cyclotron motion of the electrons in a field with a finite
component perpendicular to the plane. Naturally, this diamagnetic
response is very anisotropic and only weakly temperature dependent.
From this we can conclude that in nanographite ribbons with zigzag
edges the susceptibility should consist mainly
of these two competing contributions, $\chi_{P}$ and
$\chi_{orb.}$.  Hence, a crossover occurs from a high-temperature
diamagnetic to a low-temperature paramagnetic regime, where the
characteristic temperature depends on the width of the ribbon and of
the orientation of the external field. It is worth noting that the
field direction is an important tool to distinguish the magnitude of 
the two components.

\section{Electronic Structure of Graphite Ribbons in Magnetic Field}
\subsection{Harper Equation}
In this paper, we use a single-orbital nearest-neighbor tight binding 
model in order to study the electronic states of nanographite ribbons.
This model has been successfully used for the calculation of electronic
states of fullerene molecules, carbon nanotubes and other
carbon-related materials\cite{wallace,fujita}.
The Hamiltonian is written as,

\begin{eqnarray}
H =  \sum_{\langle i,j \rangle} t_{ij}
      c^\dagger_ic_j
\label{eq:hamil}
\end{eqnarray}

\noindent
where the operator $c^\dagger_i$  creates an electron on the site $i$,
$\langle i,j \rangle$ denotes the summation over the nearest neighbor
sites. Here, we neglect spin indices for simplicity.
In this model, the magnetic field \mbf(B) perpendicular to the
graphite plane is incorporated in the transfer integral $t_{ij}$ 
by means of the Peierls phase\cite{London} defined as

\begin{eqnarray}
 t_{ij}\longrightarrow     t_{ij}{\rm e}^{{\rm i}2\pi\phi_{i,j}},
\end{eqnarray}

\noindent
where $\phi_{i,j}$ is given by
the line integral of the vector potential from $i$-site to
$j$-site,

\begin{eqnarray}
  \phi_{i,j} = \frac{e}{ch}\int^j_i {\rm d}\mbf(l)\cdot \mbf(A).
\label{eq:phase}
\end{eqnarray}

\noindent
The magnetic flux through the area $S$ 
in units of the flux quantum $\phi_o =\frac{ch}{e}$ is

\begin{eqnarray}
  \frac{1}{\phi_0}\int{\rm d}{\bf S\cdot B}
= \frac{e}{ch}\oint{\rm d}{l\cdot A}
= \sum_{around {\bf S}}\phi_{i,j}.
\end{eqnarray}

The structure of graphite ribbons with zigzag and armchair edges are 
shown in Fig.~\ref{fig:gribbon}, where
we assume that all edge sites are terminated by hydrogen atoms.
The ribbon width
$N$  is defined by the number of zigzag lines for the 
zigzag ribbon and
by the number of dimer lines for the  armchair ribbons.
Since a hexagonal lattice can be divided by two sublattices, we call the 
A(B)-sublattice on the n-$th$ zigzag or dimer line as nA (nB) site. 
We assume Landau gauge with \mbf(A) = (0, Bx, 0), where
we define the translational invariant 
direction of each ribbon as the y-axis, and the x-axis lies
perpendicular to y-axis. In this gauge, the 
unit cell of each ribbon 
could be taken as the rectangle shown in 
Fig.~\ref{fig:gribbon}.

{\narrowtext
\begin{figure}
\hspace{-3mm}(a) \hspace{30mm}(b)\\
\epsfxsize=0.45\hsize
\epsffile{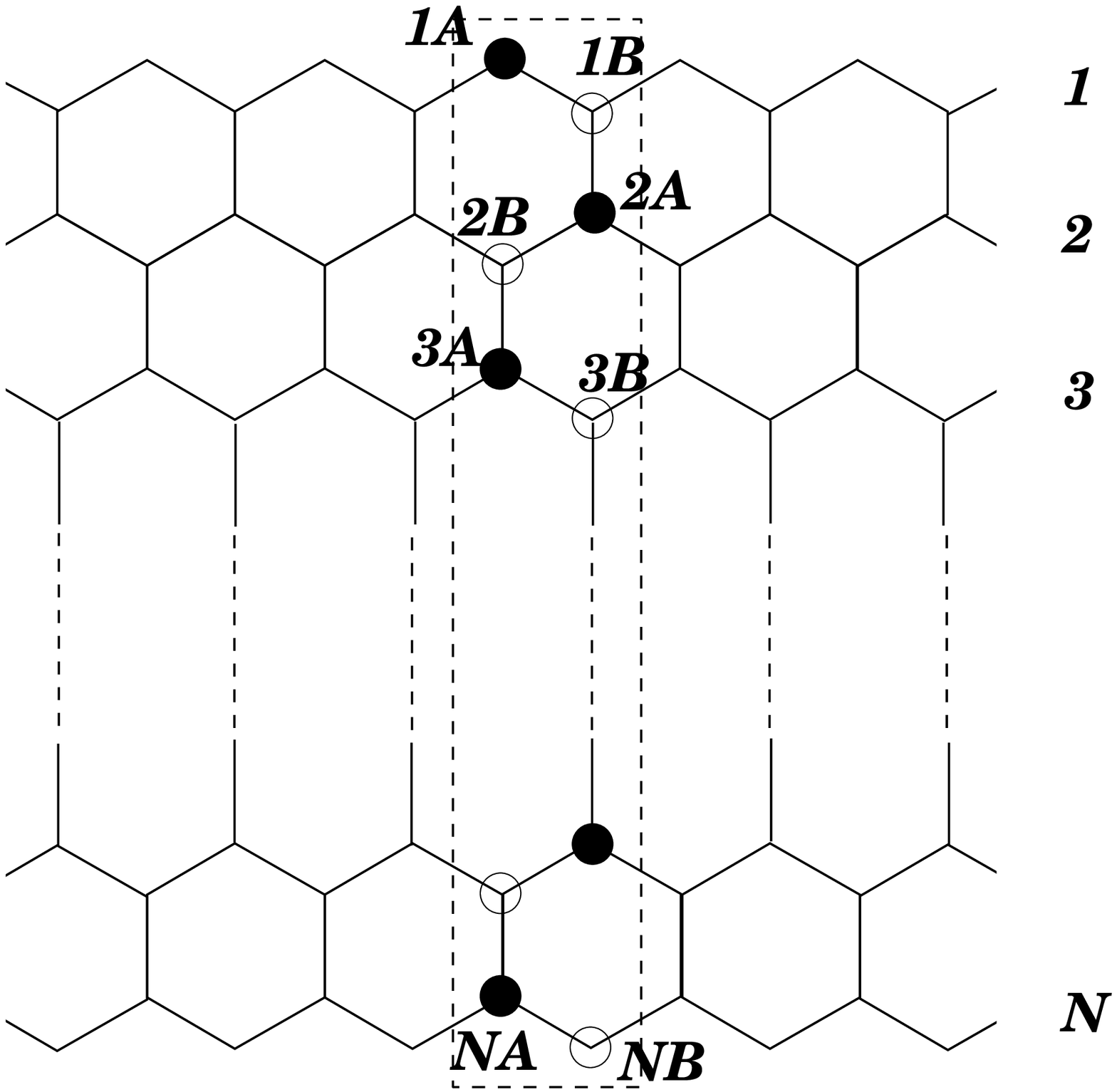}
\epsfxsize=0.5\hsize
\epsffile{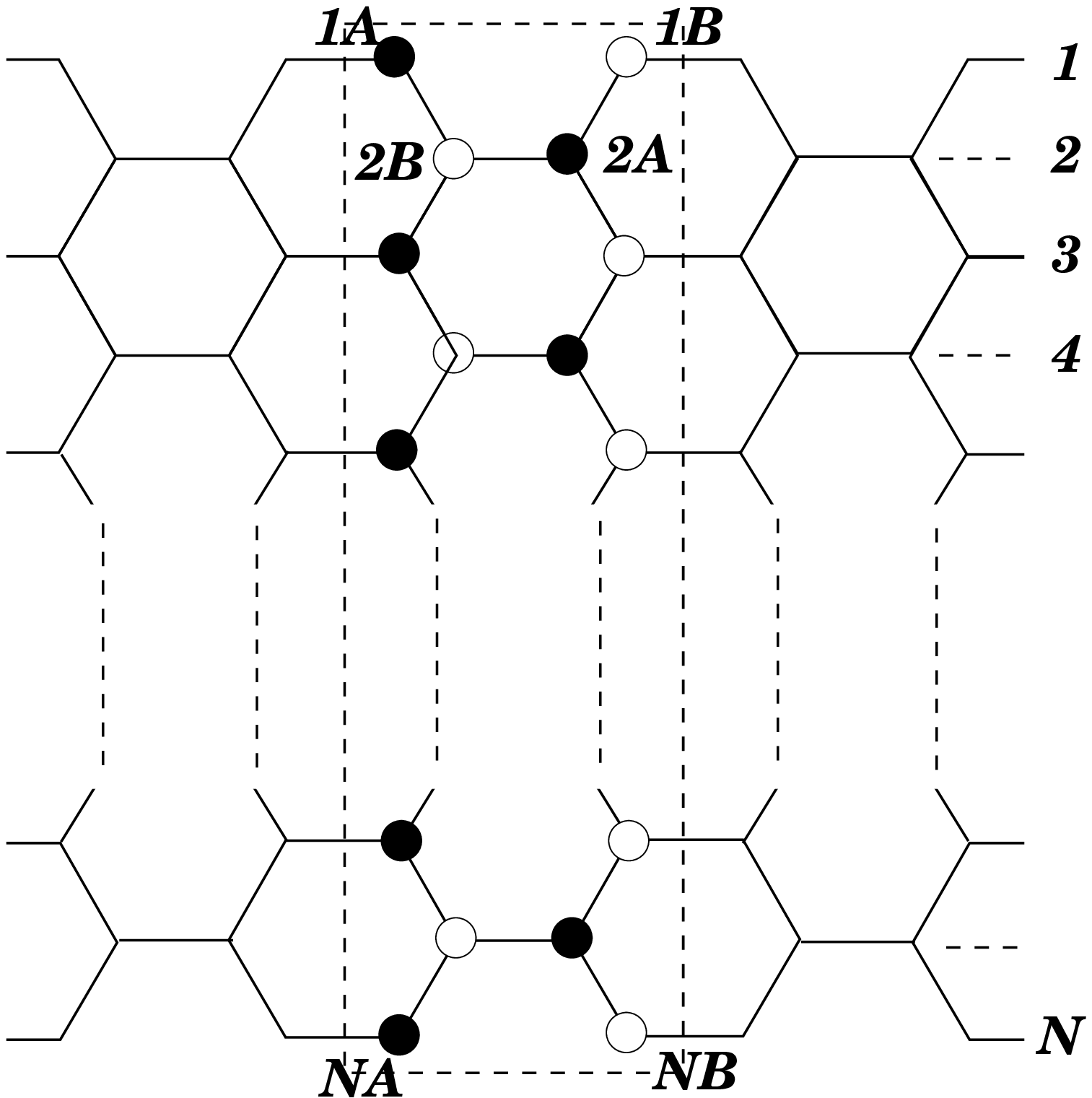}

\narrowtext
\caption{The structure of graphite ribbons with (a) zigzag edges and
(b) armchair edges. The rectangle with the dashed line is the unit
cell.}
\label{fig:gribbon}
\end{figure}
}

It should be noted that the same number  N for both zigzag
and armchair ribbons does not give the same ribbon width, when  the
ribbons are measured by the same unit of length. 
Therefore, when we 
compare physical quantities of zigzag and armchair ribbons with a
same width $W$, we will use the following definition

\begin{eqnarray}
W = \left\{
\begin{array}{ccc}
\frac{\sqrt{3}}{2}Na-a & \equiv W_z  & {\rm  zigzag \  \ ribbons}\\
(N-1)\frac{\sqrt{3}}{2}a & \equiv W_a & {\rm armchair \  \ ribbons}
\end{array}
\right.
\end{eqnarray}

\noindent
where $a$ is the C-C  bond length.

Next, let us apply Eq.(\ref{eq:hamil}) to the graphite lattice and 
derive the so called Harper equations. 
In order to apply Eq.(\ref{eq:hamil}) to the graphite lattice
and simplify the formulation,
we introduce the lattice transformation as shown in
Fig.~\ref{fig:topol},  
which does not change the lattice topology.
For convenience we will use this brick-type lattice structure when
ever we perform real calculations.

{\narrowtext
\begin{figure}
\epsfxsize=0.7\hsize
\epsffile{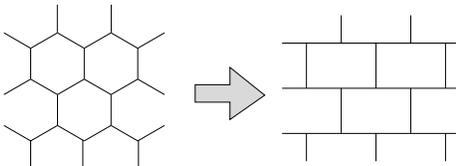}
\narrowtext
\caption{The transformation from hexagonal lattice to brick type
lattice, which does not change the lattice topology. }
\label{fig:topol}
\end{figure}
}

The Peierls phase for the graphite ribbons is easily calculated
from Eq.(\ref{eq:phase}).
The Peierls phase of graphite ribbons with
(a) zigzag and (b) armchair ribbons  is shown in 
Fig.~\ref{fig:landauzig}.
For both cases, the Peierls phase is given by
$\phi_{mB,nA}=\frac{1}{2}m\phi\delta_{mn}$, where $\phi$ is the
magnetic flux through a plaquette in units of a quantum flux.

{\narrowtext
\begin{figure}
(a)\hspace{35mm}(b)\\
\epsfxsize=0.4\hsize
\epsffile{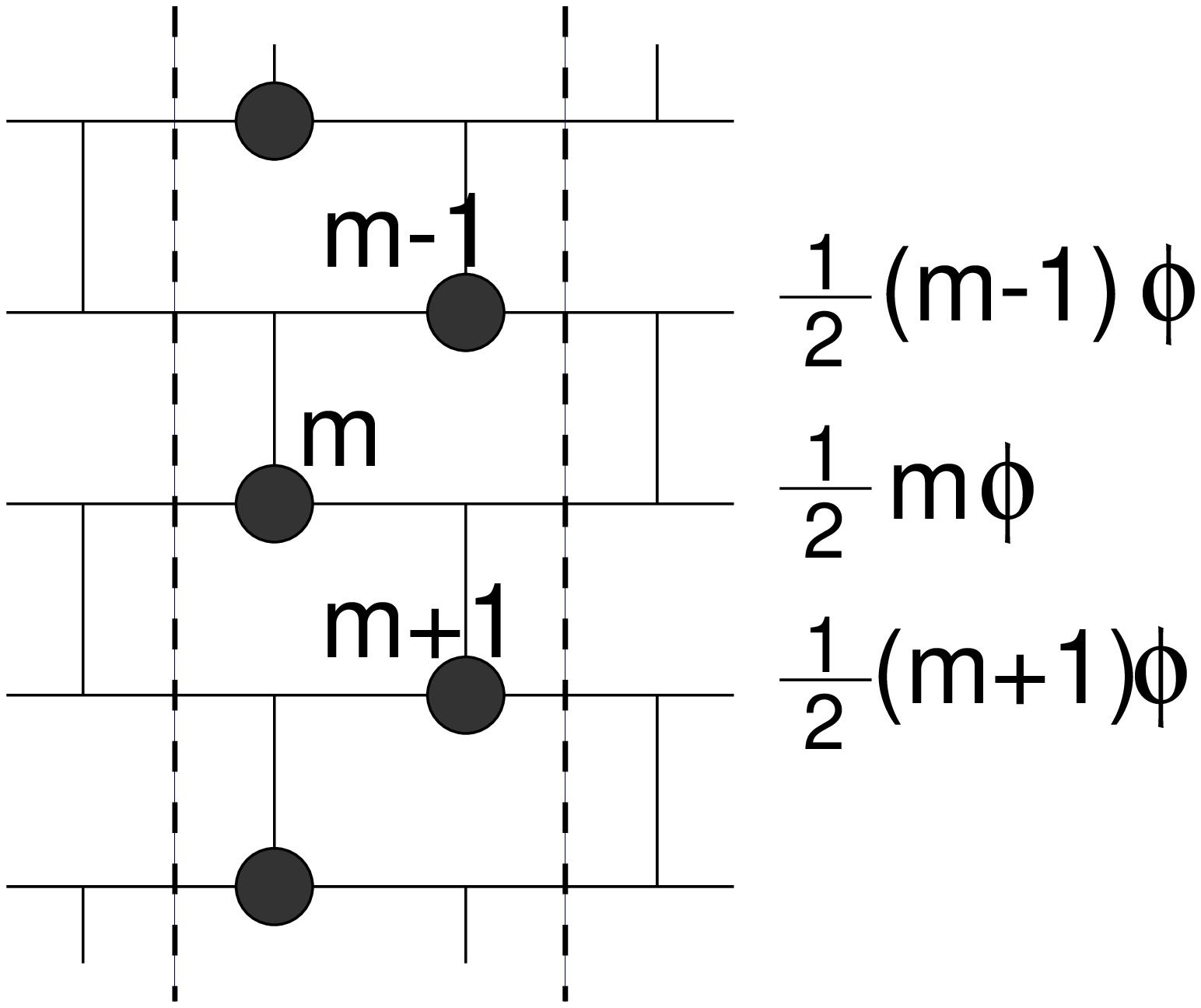}
\epsfxsize=0.48\hsize
\epsffile{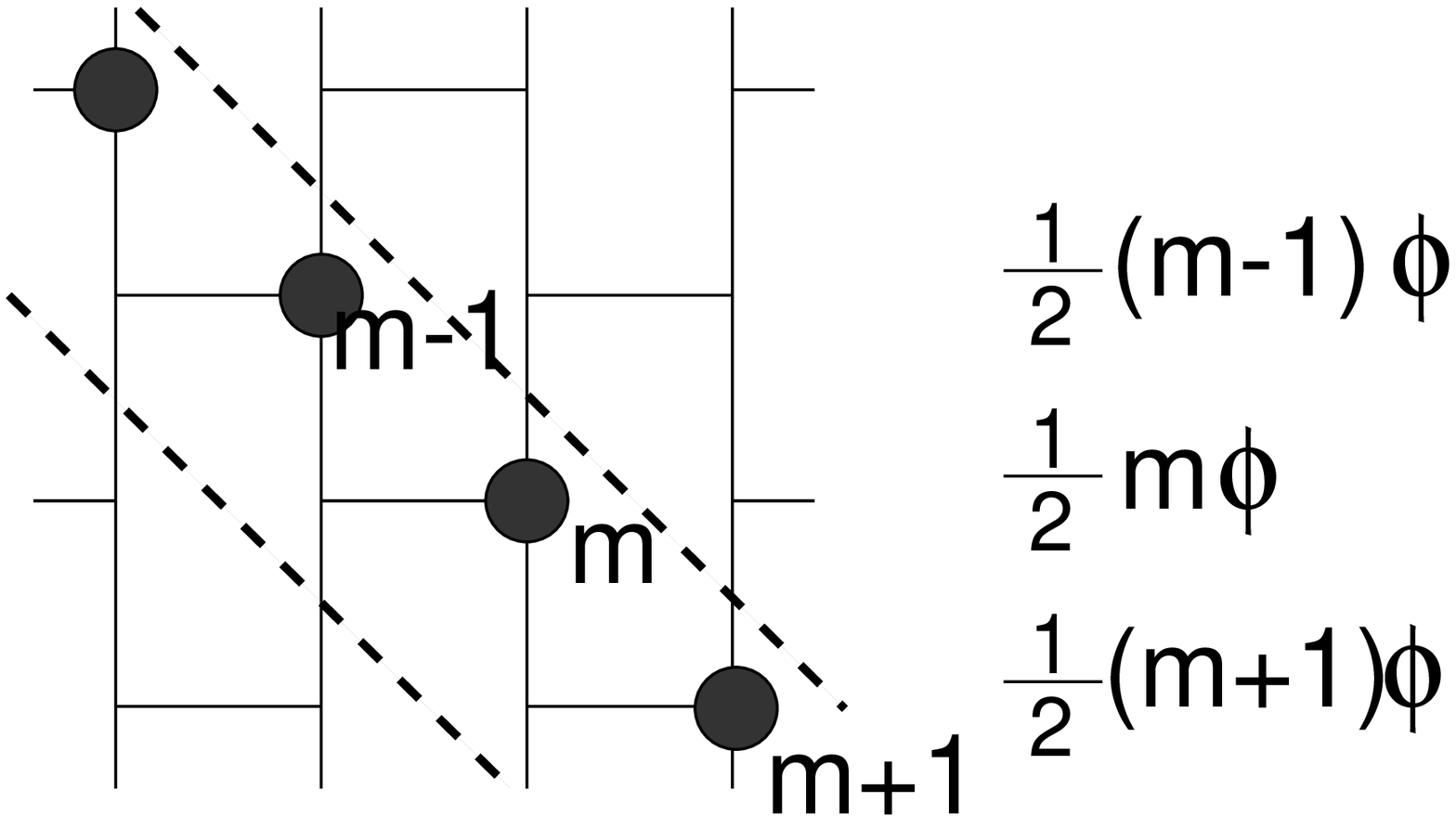}
\narrowtext
\caption{The Peierls phase for Landau gauge on (a) zigzag ribbons and 
(b) armchair ribbons.}
\label{fig:landauzig}
\end{figure}
}

Now, let us derive the Harper equation, 
which possess the translational symmetry along the 
zigzag axis.
In order to apply Eq.(\ref{eq:hamil}) to zigzag ribbons, 
we define a new operator,
$c_\alpha (i)$, which create an electron on the site i in the unit
cell $\alpha$. 
The momentum representation of this operator in the $y$-direction is
defined by

\begin{eqnarray}
  c_{\alpha}(i) = \frac{1}{\sqrt{L}}\sum_k 
  {\rm e}^{{\rm i}kr_\alpha}\gamma_{k}(i),
\end{eqnarray}

\noindent
where $r_\alpha$ represents the position of the unit cell $\alpha$.
Here we also define a one-particle state

\begin{eqnarray}
  |\Psi(k)\rangle & = & \sum_m(\Psi_{mA}(k) \gamma^\dagger_{mA}(k) 
\nonumber \\
& & + \Psi_{mB}(k) \gamma^\dagger_{mB}(k))|0\rangle .
\end{eqnarray}

\noindent
Inserting this one-particle state into  Schr\"{o}dinger equation 
$H|\Psi\rangle = \epsilon|\Psi\rangle$,
we can easily obtain the following four eigenvalue equations 
for the sites mB, mA and (m+1)A,

\begin{eqnarray}
  \epsilon\Psi_{mB} &  = & \Psi_{(m+1)A} + {\rm e}^{{\rm
 i}2\pi\frac{m}{2}\phi}\Psi_{mA} 
 + {\rm e}^{-{\rm i}2\pi\frac{m}{2}\phi}\Psi_{mA}, \nonumber \\
  \epsilon\Psi_{mA} &  = & \Psi_{(m-1)B} + {\rm e}^{{\rm
 i}2\pi\frac{m}{2}\phi}\Psi_{mB} 
 + {\rm e}^{-{\rm i}2\pi\frac{m}{2}\phi}\Psi_{mB}, \\
\label{motion}
  \epsilon\Psi_{(m+1)A} &  = & \Psi_{mB} + {\rm e}^{{\rm
 i}2\pi\frac{m}{2}\phi}\Psi_{(m+1)B} 
 + {\rm e}^{-{\rm i}2\pi\frac{m}{2}\phi}\Psi_{(m+1)B}. \nonumber
\end{eqnarray}

\noindent
Eliminating the A-sublattice sites, we obtain the difference equation,

\begin{eqnarray}
  \lambda \Psi_m(k_y) & = & a_m \Psi_{m+1}(k_y) \nonumber\\
&&+ b_m\Psi_m(k_y) 
+ a_{m-1}\Psi_{m-1}(k_y),
\label{eq:haperzig}
\end{eqnarray}

\noindent
where $\lambda=\epsilon^2-3$,
$a_m(k_y)= 2\cos(\frac{k_y}{2} + m\pi\phi)$,
$b_m(k_y)= 2\cos(k_y + 2m\pi\phi)$,
and $\Psi_{mB}$ was replaced by $\Psi_m$.
Therefore, our problem was reduced to 
a one-dimensional tight
binding model with a superlattice potential of period $2q$.
Note that this equation does not
include any boundary conditions yet. It may be applied to both
graphite ribbons and sheets by imposing the appropriate boundary
conditions. 
In the following calculations, the factor $m\pi\phi$ will be replaced
by $\left(\frac{N-1}{2}-m+1\right)\pi\phi$ to keep the energy band
symmetric about $k=0$ for arbitrary magnetic flux. 
This replacements mean that the origin of the
x-axis is set to the center of the ribbons.

The spectrum is confined to values of $\lambda$ 
between $-6$ and $+6$, i.e.
$-3\le \epsilon\le +3$. A close inspection shows that the following
translations do not 
change the energy spectrum; $\epsilon \rightarrow -\epsilon$ and $\phi
\rightarrow \phi + n$ 
,where $n$ is an arbitrary integer.
For rational flux $\phi=\frac{p}{q}$, $a_m$ is a function with
period $2q$ and $b_m$ is a function 
with period $q$. 
In addition, we must pay attention to the following symmetry of the
energy bands in the Brillouin zone, 

\begin{equation}
  \epsilon\left( k_y + \frac{2\pi}{q}n\right) =\epsilon\left( k_y
  \right). 
\end{equation}

\noindent
Here we should again note that these arguments do not depend on the
boundary conditions.

Similarly, we can also derive the Harper equations which includes
the translational symmetry along the armchair axis.
In the same way as Eq.(\ref{eq:haperzig}) the following equation is
obtained, 

\begin{eqnarray}
\lambda \Psi_m(k_y) & = & \Psi_{m+2}(k_y)+a_m \Psi_{m+1}(k_y) \nonumber \\
& &+ a_{m}\Psi_{m-1}(k_y)
+\Psi_{m-2}(k_y)
\label{eq:haperarm}
\end{eqnarray}

\noindent
where $\lambda=\epsilon^2-3$,
$a_m(k_y)= 2\cos(\frac{k_y}{2} + (m-\frac{1}{2})\pi\phi){\rm
e}^{-{\rm i}\frac{\pi}{2}\phi}$. 
It is easy to confirm that the same symmetry properties as for
Eq.(\ref{eq:haperzig}) apply here too.
Similarly, the factor $a_m$  will be replaced by
$a_m(k_y)= 2\cos(\frac{k_y}{2} + (\frac{N-1}{2}-2m+3)\pi\phi){\rm
e}^{-{\rm i}\pi\phi}$ in order to obtain the symmetric energy band
structure  about $k= 0$ for arbitrary magnetic flux.

\subsection{Graphite Sheet}
In this subsection, we consider the electronic structures of a graphite 
sheet in a magnetic field. This will afterwards become the basis to
discuss the electronic structures of nanographite ribbons.
As we have seen in the previous subsection,
the tight binding model of the graphite lattice could be reduced to a
tight binding model with superlattice potential of period
$2q$. In order to calculate the energy spectrum of the graphite sheet,  
we must treat the eigenvalue problem of a $2q\times2q$ matrix with the 
periodic boundary condition 
$\Psi_{2q+1} = {\rm e}^{{\rm i}k_x 2q}\Psi_1$, 
when the Brillouin zone is reduced to the magnetic Brillouin zone
$-\frac{\pi}{2q}\le k_x \le \frac{\pi}{2q}$ and  
$-\pi \le k_y \le \pi$.

At the beginning, we consider the zero-field energy band of the
graphite sheet. By setting $\phi =0$ in Eqs.(\ref{eq:haperzig}) and
(\ref{eq:haperarm}), the zero-field spectrum is easily obtained.
From Eq.(\ref{eq:haperzig}), 
we find for the graphite sheet with translational
symmetry along zigzag axis,

\begin{eqnarray}
  \epsilon_{\mbf(k)} = \pm \sqrt{3+ 2\cos(k_y) +
  4\cos\left(\frac{k_y}{2}\right) \cos(k_x)}. 
\label{eq:gsheet_zig}
\end{eqnarray}

Similarly, from Eq.(\ref{eq:haperarm}),
the zero-field spectrum of graphites with translational
symmetry along the armchair axis is

\begin{eqnarray}
  \epsilon_{\mbf(k)} = \pm \sqrt{3+ 2\cos(2k_x) +
4\cos\left(\frac{k_y}{2}\right)\cos(k_x)}. 
\label{eq:gsheet_arm}
\end{eqnarray}

The energy band structures for both  Eqs.(\ref{eq:gsheet_zig}) and
(\ref{eq:gsheet_arm}) are shown in Fig.~\ref{fig:2dgraband}, 
where $k_y$ is replaced by $k$ and we have superposed all $k_x$-values
in the spectrum.  We can find the degeneracy at $\epsilon = 0$ in both
figures, which originates from 
the K-point degeneracy of the band structure of the
graphite sheet\cite{peculiar,nakada}.

{\narrowtext
\begin{figure}
(a)\hspace{30mm}(b)\\
\hspace{5mm}
\epsfxsize=0.35\hsize
\epsffile{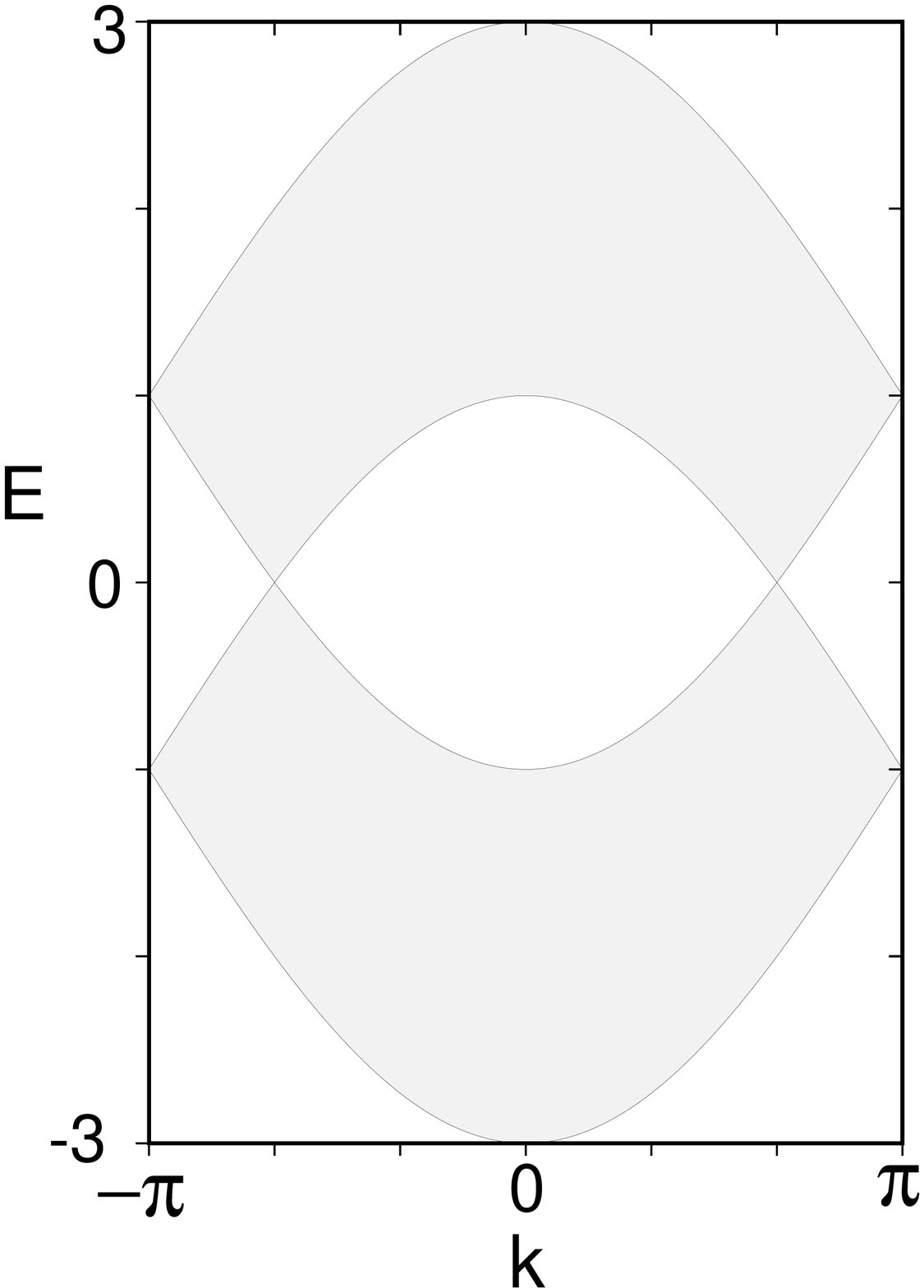}
\hspace{5mm}
\epsfxsize=0.35\hsize
\epsffile{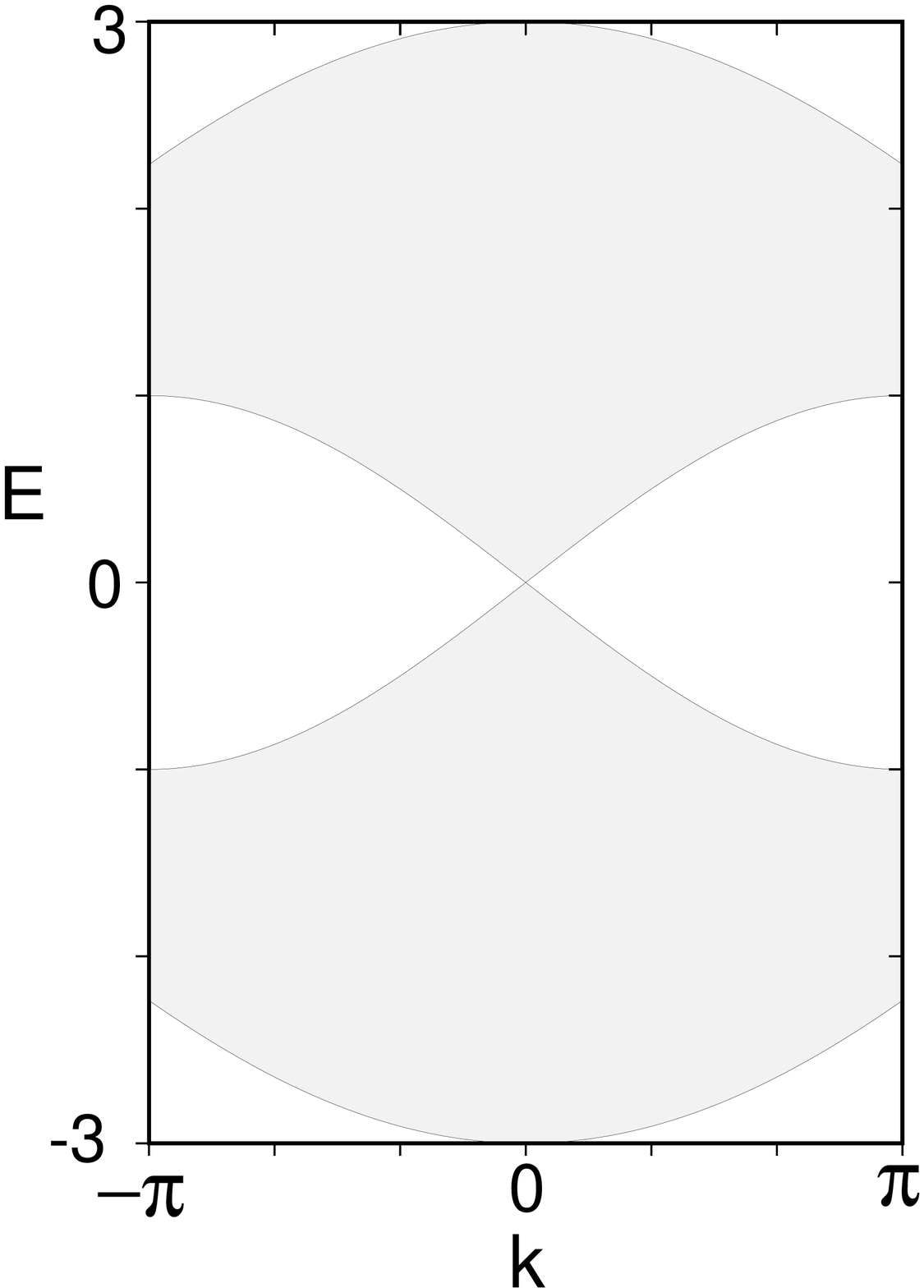}
\narrowtext
\caption{The energy band structure of graphite sheet projected to (a)
zigzag axis and (b) armchair axis in the absence of magnetic field.}
\label{fig:2dgraband}
\end{figure}
}

The spectrum of the graphite sheet in a magnetic field
is shown in Fig.~\ref{fig:2dgraspect}, which was first calculated
by Rammal\cite{rammal}.
We can easily find that the spectrum has the fine recursive
structure of the Hofstadter butterfly. 
In the weak-magnetic field limit, we can clearly see the Landau levels.
When the magnetic flux is getting larger, 
these levels form the
Landau subbands because of the Harper broadening.
As we pointed out in the previous subsection, for the rational flux
$\phi=\frac{p}{q}$, we can see 
$2q$ subbands with a reflection symmetry about $\epsilon =0$ and about 
$\phi=\frac{1}{2}$. 
Interestingly, the degeneracy at $\epsilon=0$ exits for arbitrary
flux, which confirms that the K-point degeneracy  
will not be destroyed.

{\narrowtext
\begin{figure}
\epsfxsize=0.8\hsize
\epsffile{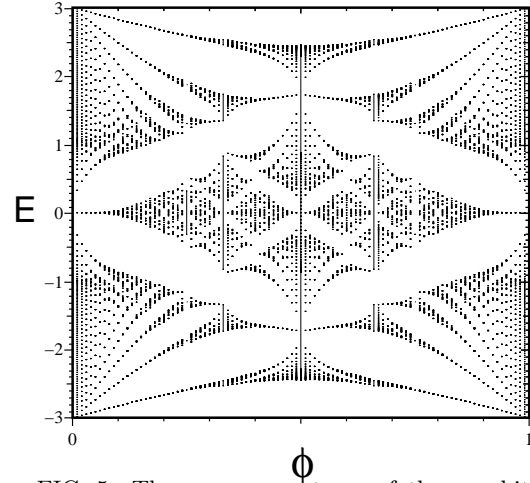}
\caption{The energy spectrum of the graphite sheet in a magnetic field.}
\label{fig:2dgraspect}
\end{figure}
}

{\narrowtext
\begin{figure}
(a) \hspace{30mm}(b)\\
\hspace{5mm}
\epsfxsize=0.35\hsize
\epsffile{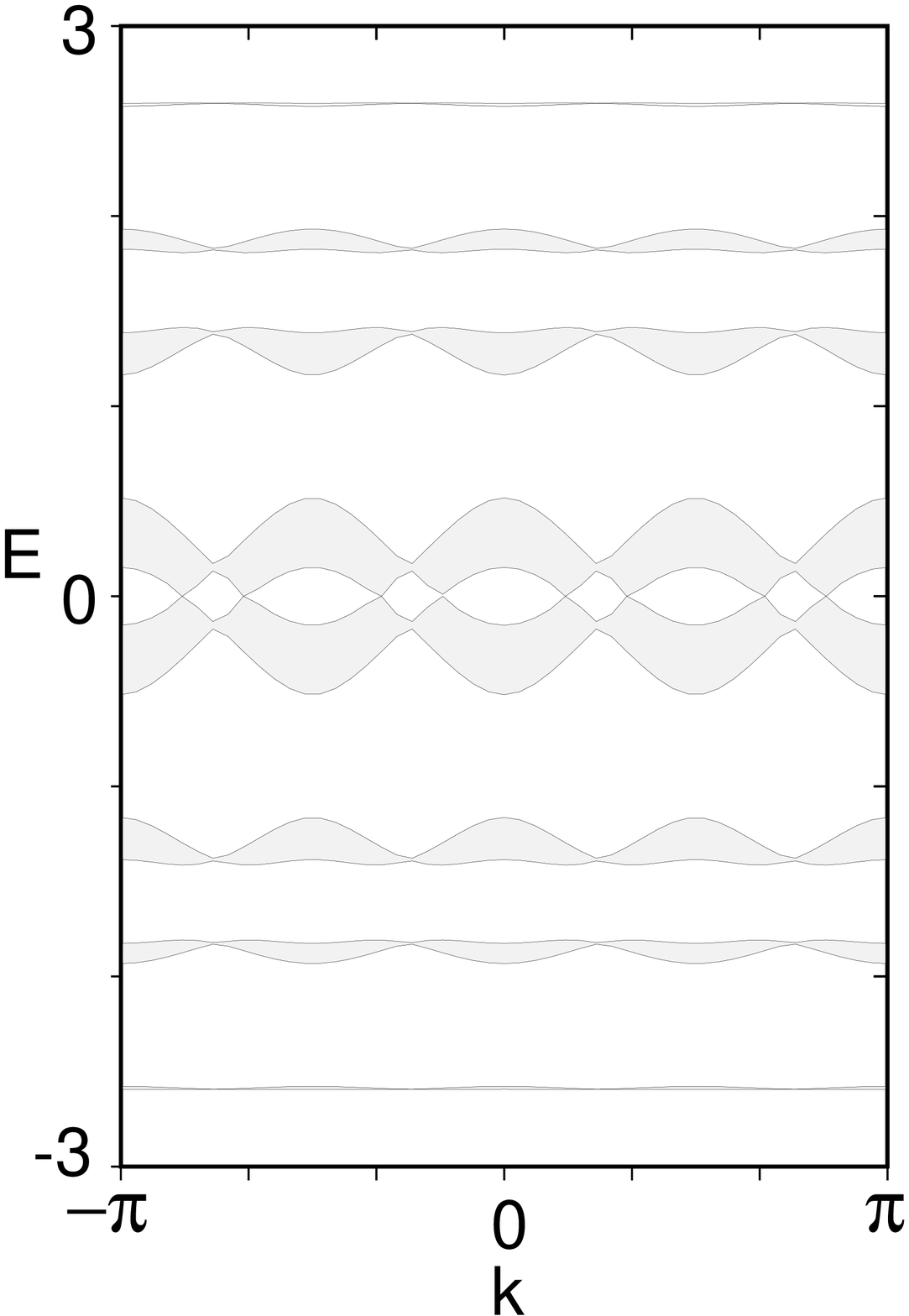}
\hspace{5mm}
\epsfxsize=0.35\hsize
\epsffile{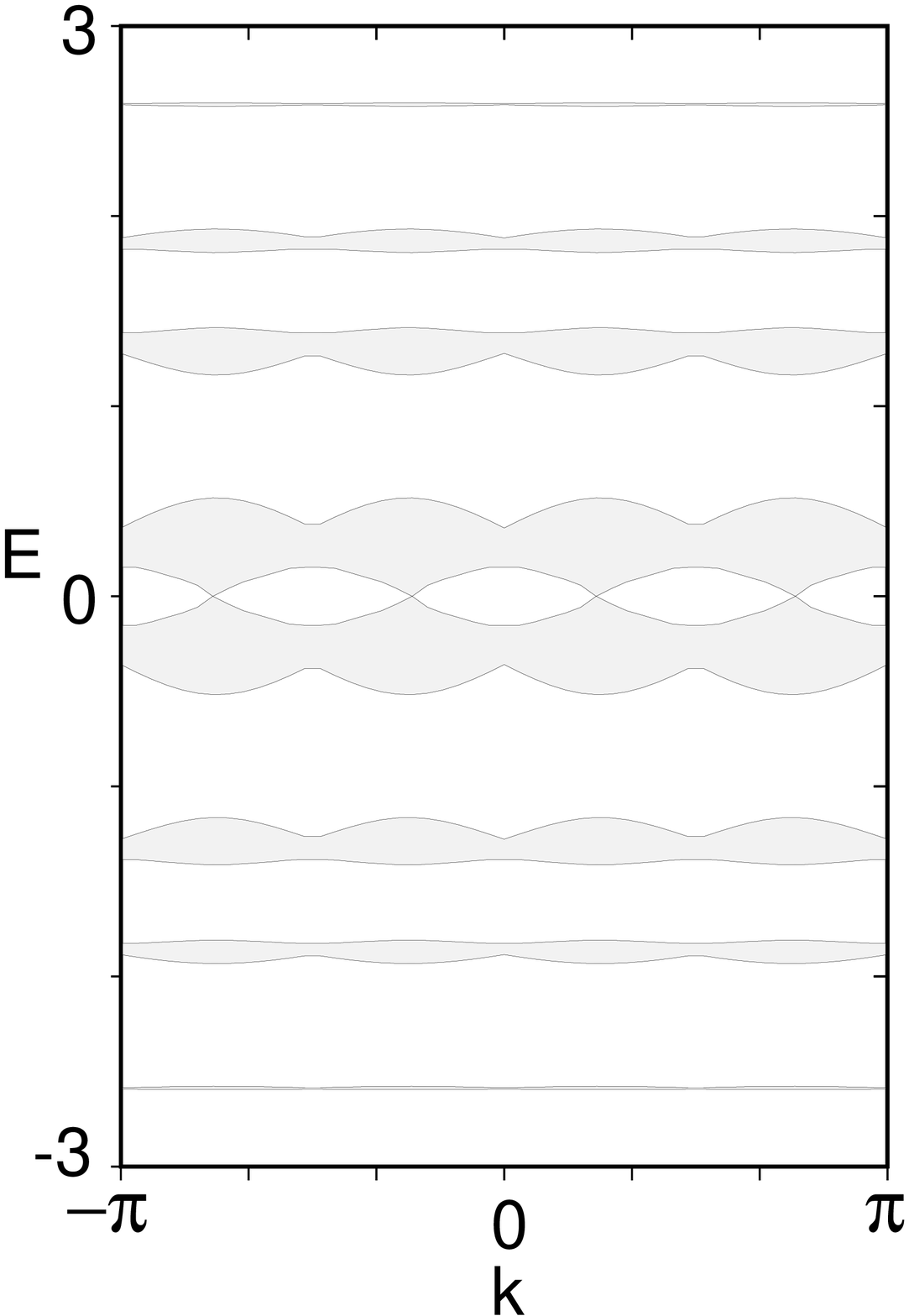}
\caption{The energy band structure of graphite sheet projected to (a)
zigzag axis and (b) 
armchair axis for $\phi=1/4$.}
\label{fig:gsheet_landau_arm}
\end{figure}
}

As an example of the energy band structure in a strong magnetic field,
we show the case of the graphite sheet projected to the (a) zigzag
axis and (b) armchair axis for $\phi=1/4$
(Fig.~\ref{fig:gsheet_landau_arm}). 
In a strong magnetic field, the Landau levels
change to $2\times 4$ subbands due to Harper broadening and the
effect of the lattice structure gets more important, so that 
each subband has the basic structure of the  zero field case.
Furthermore we find the symmetry 
$\epsilon\left( k_y + \frac{2\pi}{q}n\right) =\epsilon\left( k_y
\right)$.

\subsection{Graphite Ribbon}
The energy band structures of graphite ribbons are obtained in terms
of the Harper equation by imposing open boundary conditions.
In the case of zigzag ribbons with width $N$, the boundary condition is 
$\Psi_{N+1} =\Psi_0 = 0$. However, we need a more careful treatment of
the Harper equation at the edge site. In Eqs.(\ref{motion}), it was
not considered that there would be no 0A(B) and (N+1)A(B) site.
Including this fact, only for $m=1$ and $N$, 
the Eq.(\ref{eq:haperzig}) has to be  rewritten as

\begin{eqnarray}
  \lambda \Psi_m(k_y) & = & a_m \Psi_{m+1}(k_y) \nonumber\\
&&+ (b_m-1)\Psi_m(k_y) =
+ a_{m-1}\Psi_{m-1}(k_y).
\label{eq:haperzig_edge}
\end{eqnarray}

\noindent
Therefore we must replace $b_1$ ($b_N$) by $b_1
-1 $ ($b_N-1$) in Eq.(\ref{eq:haperzig})  in order to include the 
condition that the 0A and (N+1)A sites do not exist.
Similarly, for the armchair ribbons, the boundary condition is 
$\Psi_{N+1} =\Psi_0 = 0$, where we must replace $b_1$ ($b_N$) by $b_1
-1 $ ($b_N-1$)  in order to include the 
condition that the 0A and (N+1)A sites do not exist.

We show the energy band structures of the zigzag ribbon with $N=50$
for $\phi=0, \frac{1}{500},\frac{1}{100},\frac{1}{4}$ in
Fig.~\ref{fig:disp_zig} (a)-(d).  
For $\phi=0$, the profile of band structure has almost the same
structure as in the case of the 
graphite sheet as shown in Fig.~\ref{fig:2dgraband} (a).
However, we can see partly flat bands at $E=0$, which do not appear
in the energy band of the graphite sheet. 
The electronic states corresponding to partly flat bands are
the strongly localized states near the zigzag edges, called
``edge states''.
Analytical properties will be discussed in the next section.

In the case of $q \geq N$,
the Landau levels are not perfectly formed  
and the band structure at $\phi=0$ is almost unchanged,
because the edges interrupt the cyclotron motion of the electron
(Fig.~\ref{fig:disp_zig} (b)).
For $q\leq N$,  where the ribbon width is
sufficiently wide compared with the cyclotron radius,
the Landau levels are nearly developed (Fig.~\ref{fig:disp_zig} (c)).

As an example for $q\ll N$ and higher commensurates,
we show the energy band structure of $\phi=\frac{1}{4}$ in
Fig.~\ref{fig:disp_zig} (d).
The $2\times4$-Landau subbands are formed.
Between the Landau subbands, we can see the additional dispersion.
The states of these dispersions are also 
localized at the edge\cite{macdo,hatsugai1,rammal2}, 
which 
originate from the cyclotron motion of electrons and not 
for topological
reason. It should be noted that the partly flat bands
are formed at $E=0$ again.

{\narrowtext
\begin{figure}[t]
(a) \hspace{30mm}(b)\\
\hspace{5mm}
\epsfxsize=0.35\hsize
\epsffile{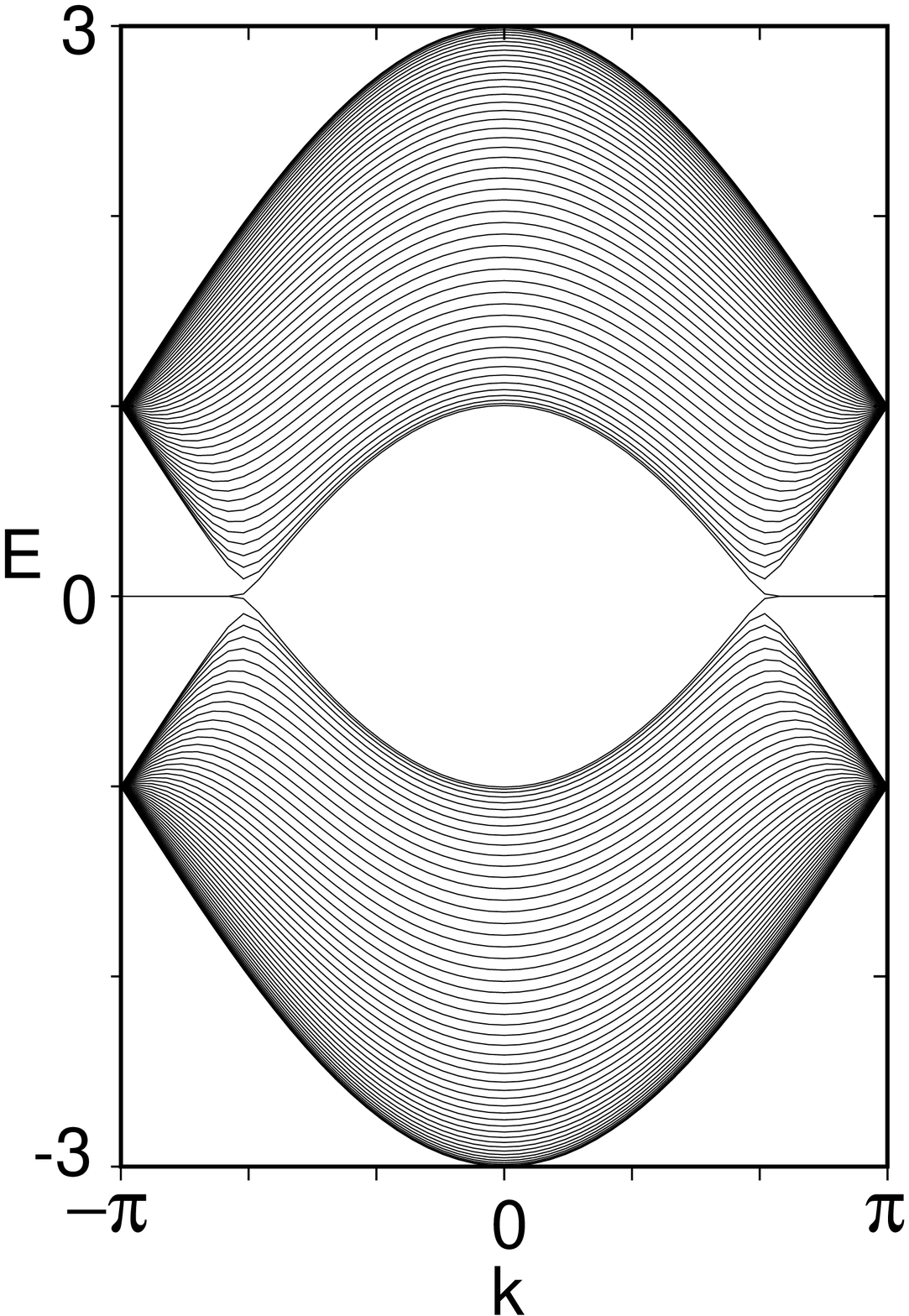}
\epsfxsize=0.35\hsize
\epsffile{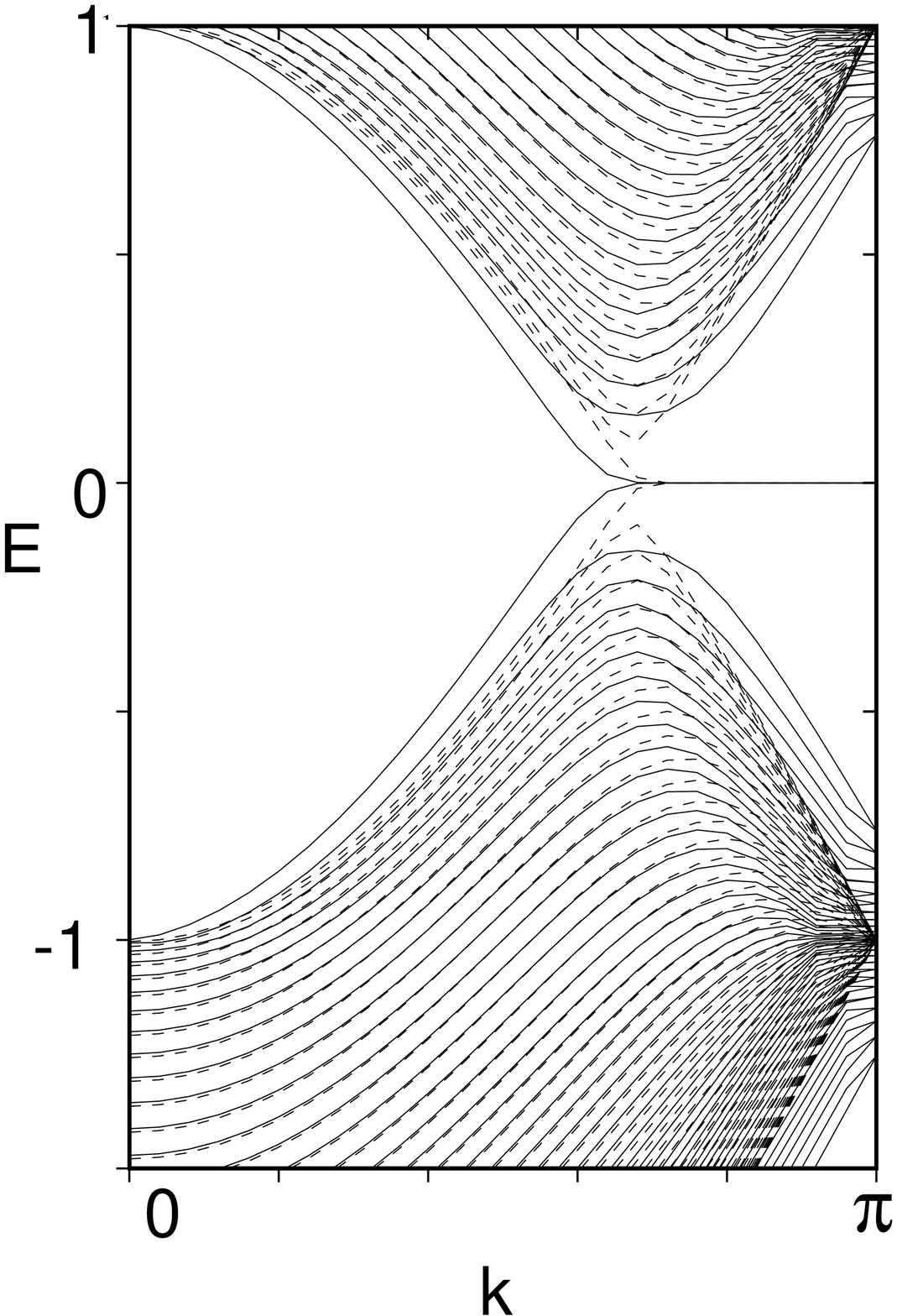}\\
(c) \hspace{30mm} (d)\\
\hspace{5mm}
\epsfxsize=0.35\hsize
\epsffile{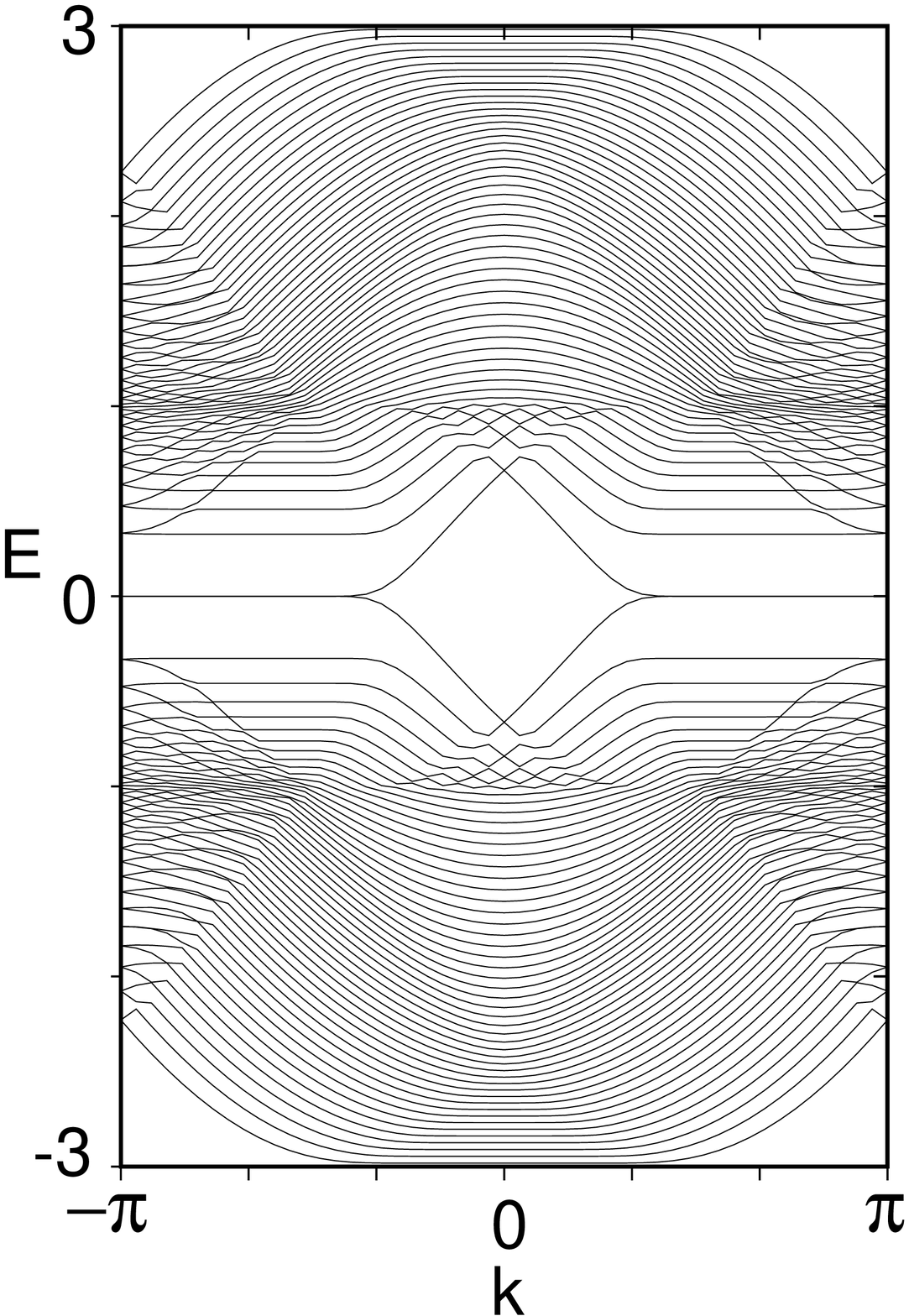}
\epsfxsize=0.35\hsize
\epsffile{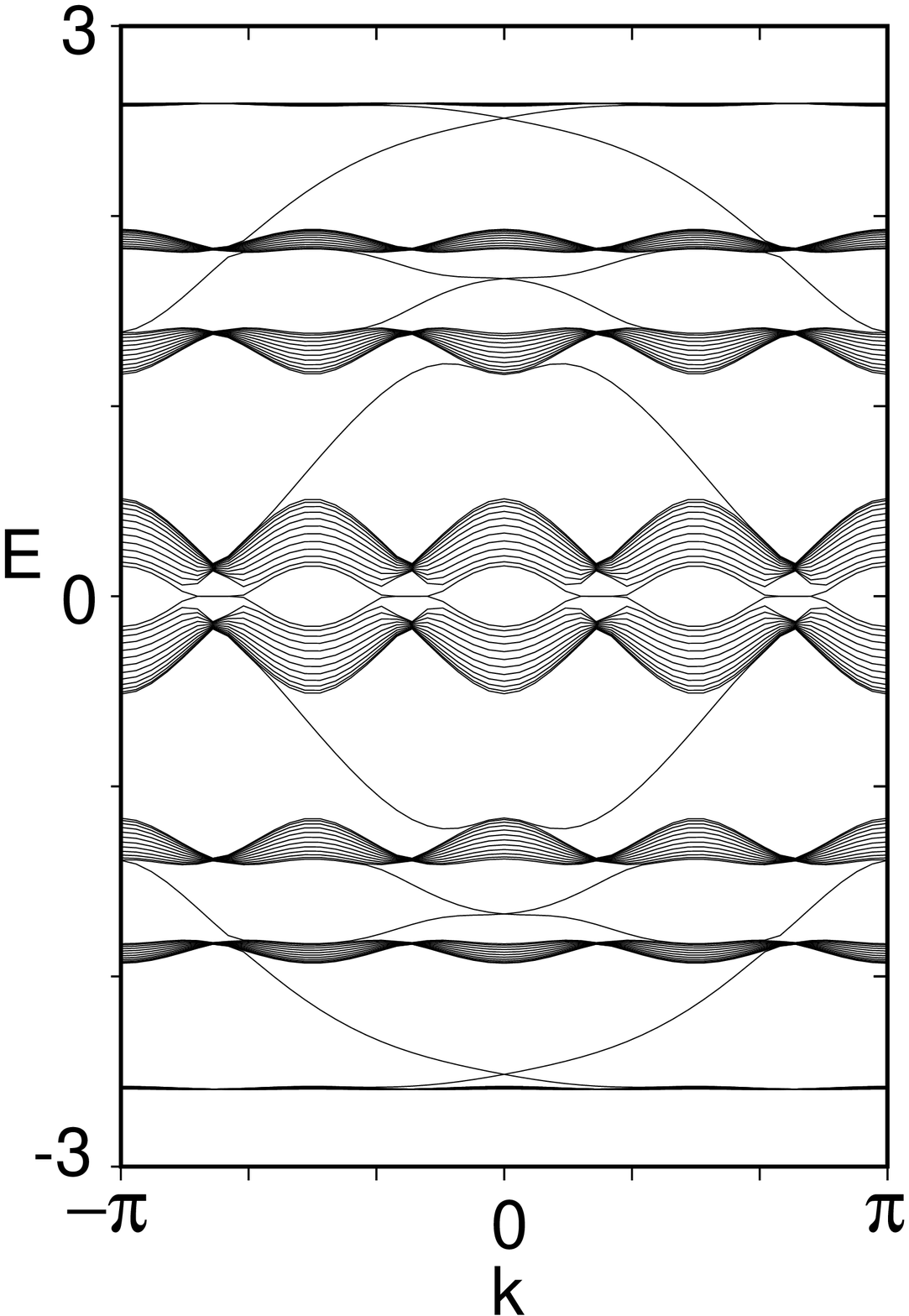}
\caption{The energy band structures of zigzag ribbon with $N=50$ for
(a) $\phi=0$, (b) $\phi=\frac{1}{500}$ 
, (c) $\phi=\frac{1}{100}$ and (d) $\phi =\frac{1}{4}$ for
$\phi=\frac{1}{4}$. 
The dashed line in (b) is the energy structure at
$\phi = 0$ for comparison.}
\label{fig:disp_zig}
\end{figure}
}

{\narrowtext
\begin{figure}
(a) \hspace{30mm}(b)\\
\hspace{5mm}
\epsfxsize=0.35\hsize
\epsffile{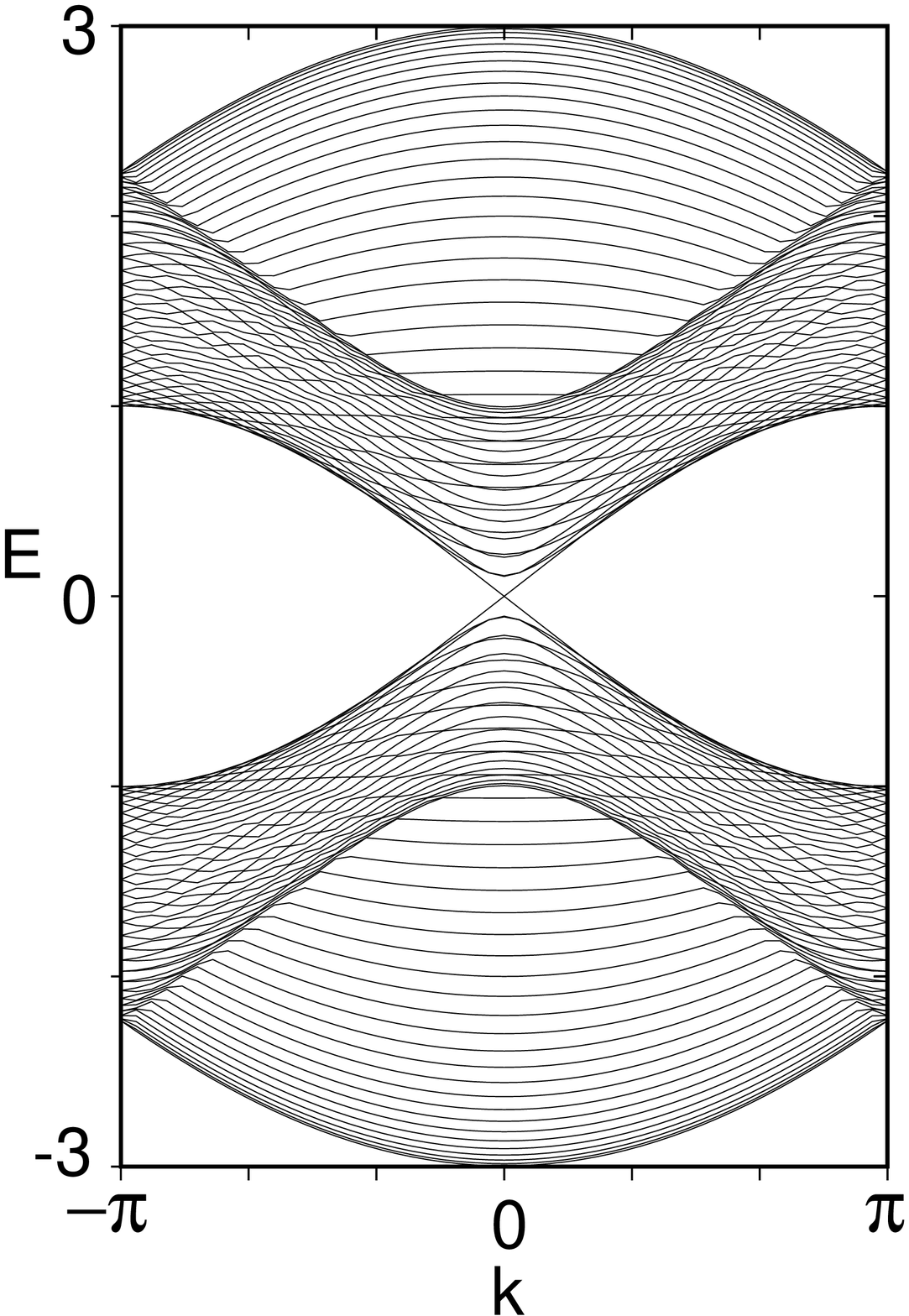}
\epsfxsize=0.35\hsize
\epsffile{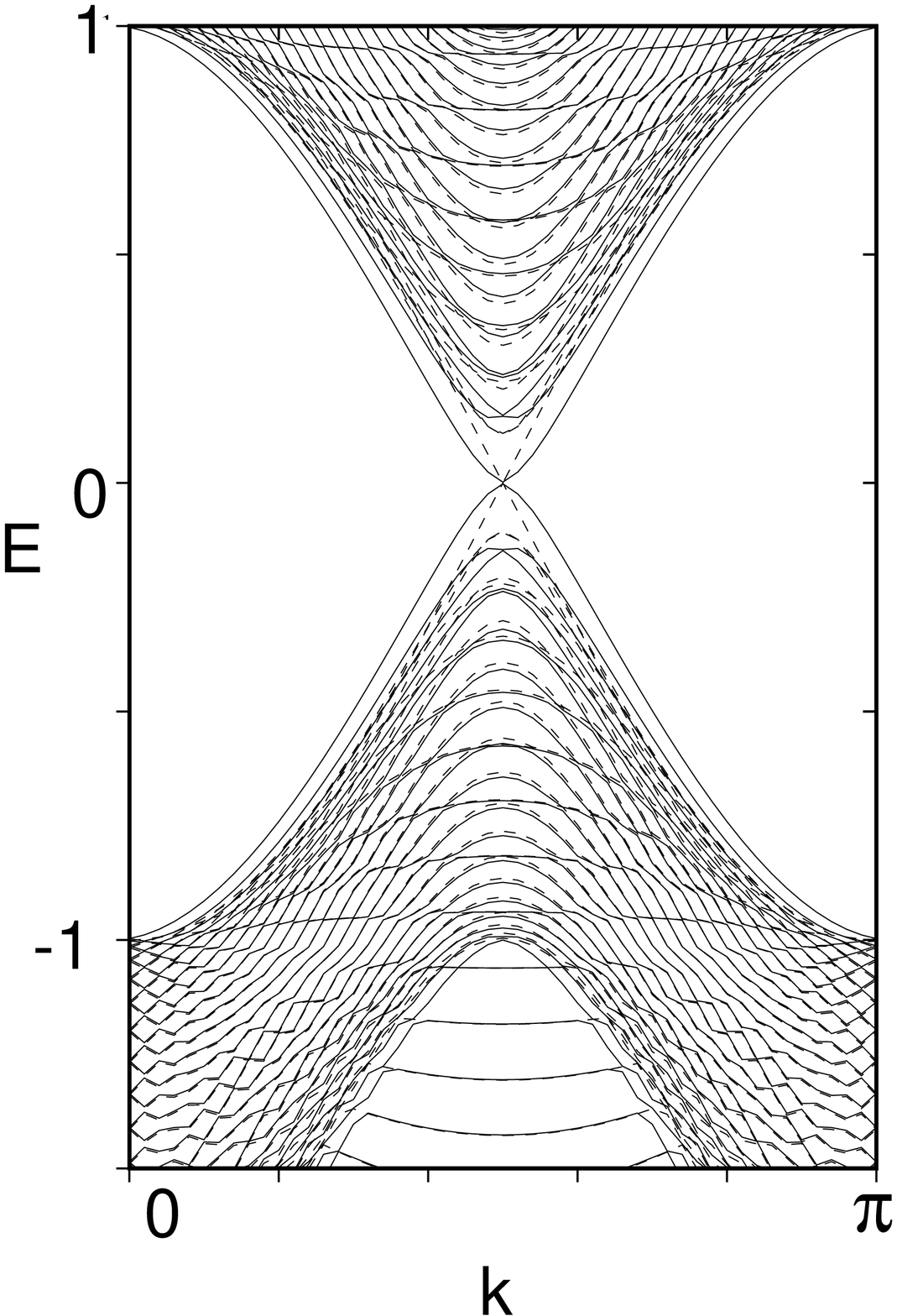}\\
(a) \hspace{30mm}(b)\\
\hspace{5mm}
\epsfxsize=0.35\hsize
\epsffile{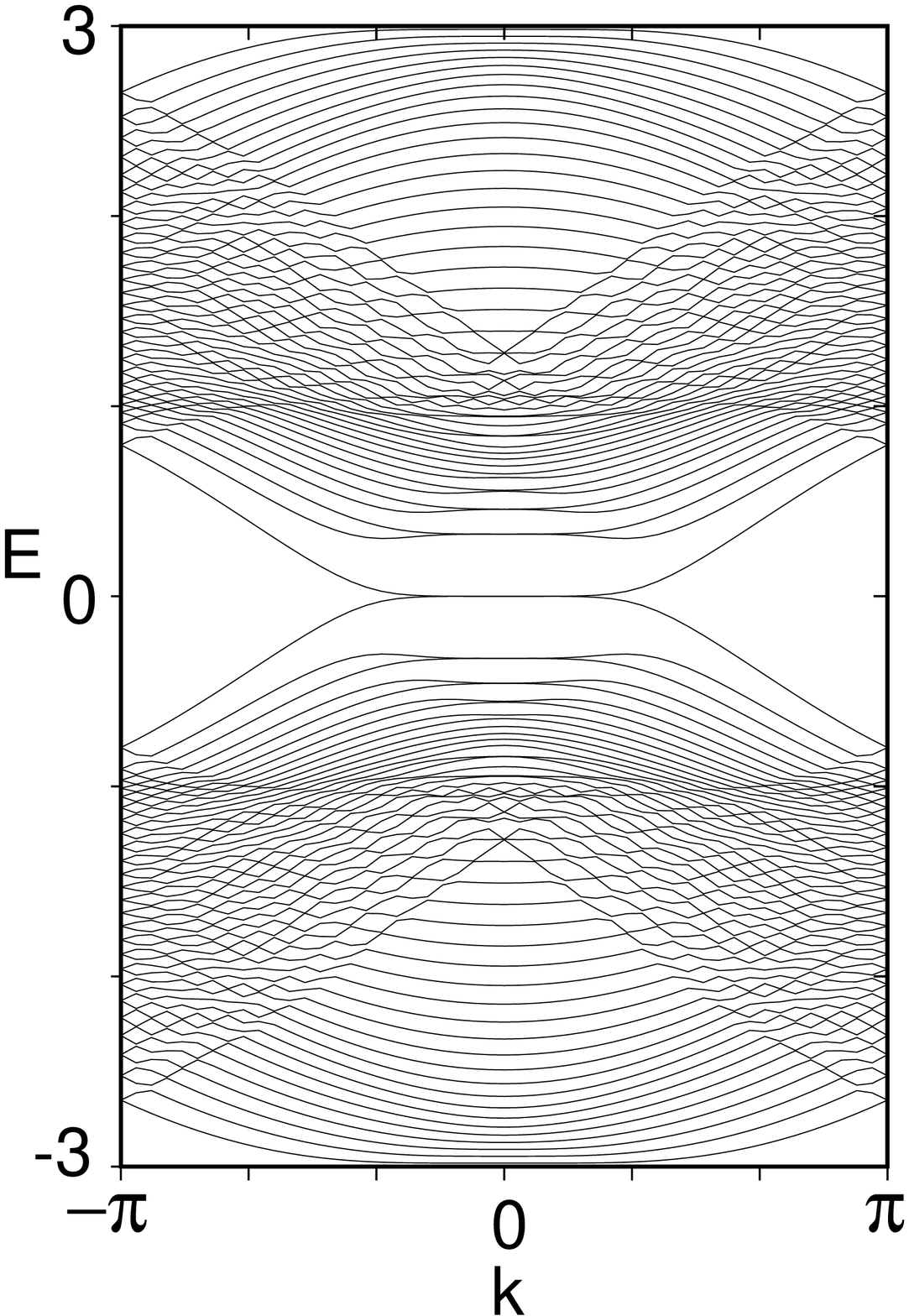}
\epsfxsize=0.35\hsize
\epsffile{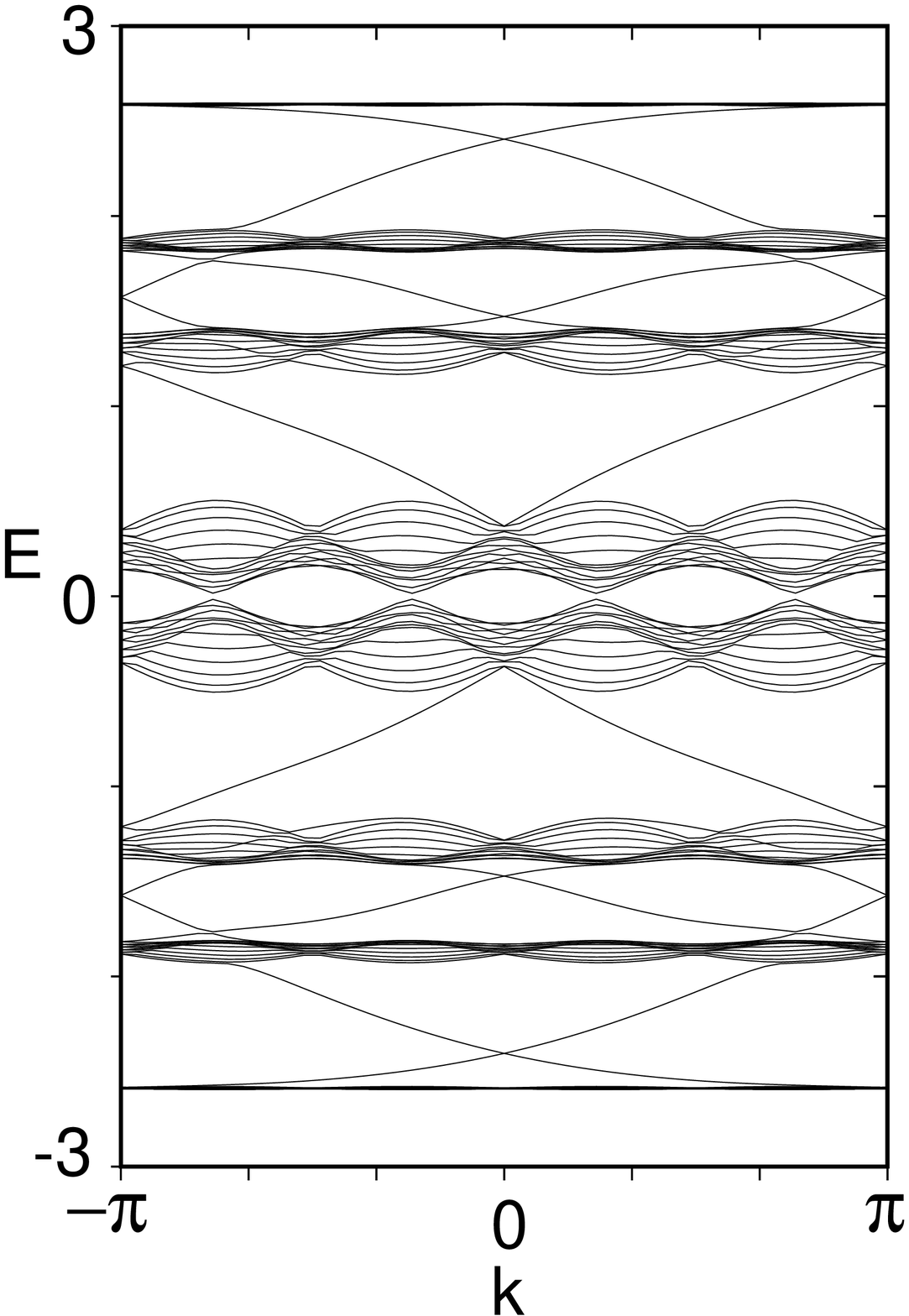}
\caption{The energy band structures of armchair ribbon with $N=50$ for
(a) $\phi=0$, (b) $\phi=\frac{1}{500}$ , (c) $\phi=\frac{1}{100}$ 
and
(d) $\phi =\frac{1}{4}$ for $\phi=\frac{1}{4}$.
The dashed line in (b) is the energy structure at
$\phi = 0$ for comparison.}
\label{fig:disp_arm}
\end{figure}
}

Next, we show the energy band structures of the armchair ribbon with
$N=50$ for $\phi=0, \frac{1}{500},\frac{1}{100},\frac{1}{4}$ in
Fig.~\ref{fig:disp_arm} (a)-(d).   For $\phi=0$, the profile of band
structure is almost identical to the one of the graphite sheet
( Fig.~\ref{fig:2dgraspect} (a)). Here, we can not see partly flat
bands at $E=0$. 

In the case of $q \geq N$,  the 
Landau levels are not perfectly formed  
and the band structure at $\phi=0$ does not change.
For $q\leq N$,  where the ribbon width is
sufficiently wide compared with the cyclotron radius,
the Landau levels are again almost formed as shown in
Fig.~\ref{fig:disp_arm}.
For the case of $q\ll N$ and higher commensurate fluxes,
we show the energy bands structure of $\phi=\frac{1}{4}$ in
Fig.~\ref{fig:disp_arm} (d).
We also find that despite the additional dispersion between Landau
subbands, no partly flat bands, as may occur in zigzag ribbons,
are present at $E=0$ even in a magnetic field.

\section{Edge State}
The states corresponding to the partly flat bands are analytically
derived by Fujita  and coworkers for the semi-infinite graphite sheet 
with a zigzag edge\cite{peculiar,nakada}. 
It can be understood as localized states near the 
zigzag edge.  It is also possible to find the edge states
by solving the Harper equation (\ref{eq:haperzig}).

At the beginning, let us rewrite the Eq.(\ref{eq:haperzig}) to the 
transfer matrix form,

\begin{eqnarray}
 \left(
  \begin{array}{l}
    \Psi_{m+1} \\
    \Psi_{m}
  \end{array}
 \right)
=
 \left(
  \begin{array}{cc}
    \frac{1}{a_m}\left(\lambda-\tilde{b}_m\right) &
    \frac{a_{m-1}}{a_m} \\ 
    1 & 0 

  \end{array}
 \right)
 \left(
  \begin{array}{c}
    \Psi_m \\
    \Psi_{m-1}
  \end{array}
 \right),
\label{eq:tmatrix}
\end{eqnarray}

\noindent
where $\tilde{b}_m$  
is $b_m -1$ for $m=1,N$ and $b_m$ for others.  
Let us take open boundary conditions, i.e
\begin{eqnarray}
 \left(
  \begin{array}{c}
    \Psi_1 \\
    \Psi_0
  \end{array}
 \right)
=
 \left(
  \begin{array}{c}
    1 \\
    0
  \end{array}
 \right) 
\label{eq:init}
\end{eqnarray}
\noindent
and we impose the condition $\epsilon = 0$.
One can obtain,

\begin{eqnarray}
  \Psi_n =D_k^{n-1},
\label{eq:estate}
\end{eqnarray}

\noindent
where $D_k$  is  $-2\cos(\frac{k}{2})$. Then the convergence
condition $| D_k | \le 1$ is required, 
because otherwise the wave function would diverge in the semi-infinite
graphite sheet.
This convergence condition defines the region $\frac{2}{3}\pi\le
k\le\pi$, where the partly flat bands exist. 
The charge density is shown in Fig.~\ref{fig:charge} at (a) $k=\pi$, 
(b) $\pi$, (c) $\pi$ and (d) $\frac{2}{3}\pi$.
At $k=\pi$, the charge is perfectly localized at the zigzag edge.
When the wave number deviates from $k=\pi$, the electron gradually
penetrates towards the inner sites. Finally, the electron states
completely extend at $k=\frac{2}{3}\pi$.

{\narrowtext
\begin{figure}
(a) \hspace{30mm}(b)\\
\hspace{5mm}
\epsfxsize=0.35\hsize
\epsffile{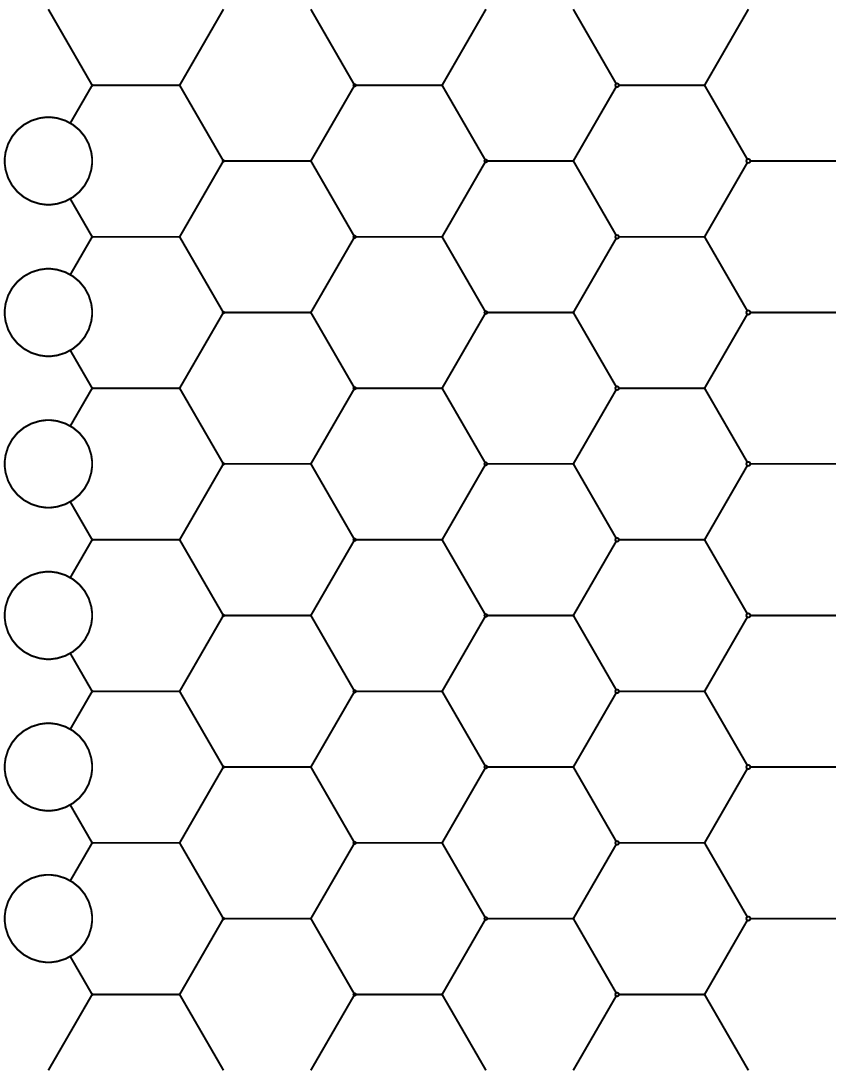}
\hspace{5mm}
\epsfxsize=0.35\hsize
\epsffile{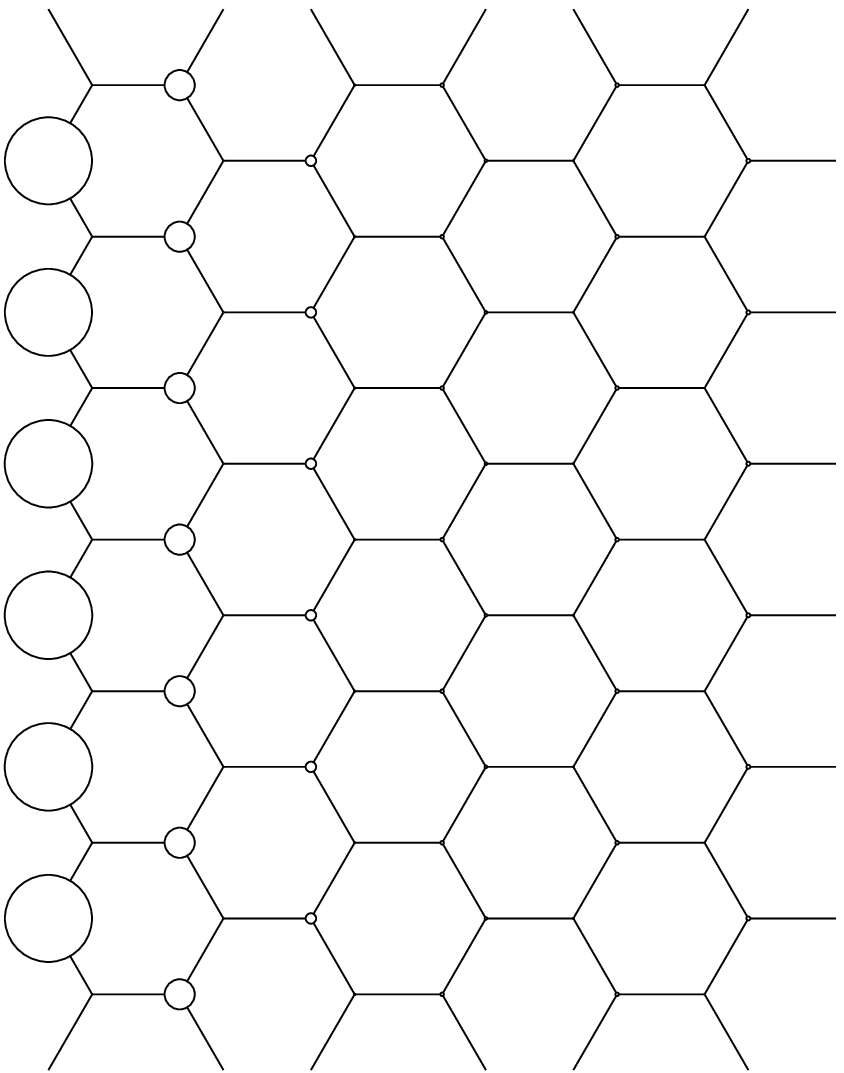}\\
(a) \hspace{30mm}(b)\\
\hspace{5mm}
\epsfxsize=0.35\hsize
\epsffile{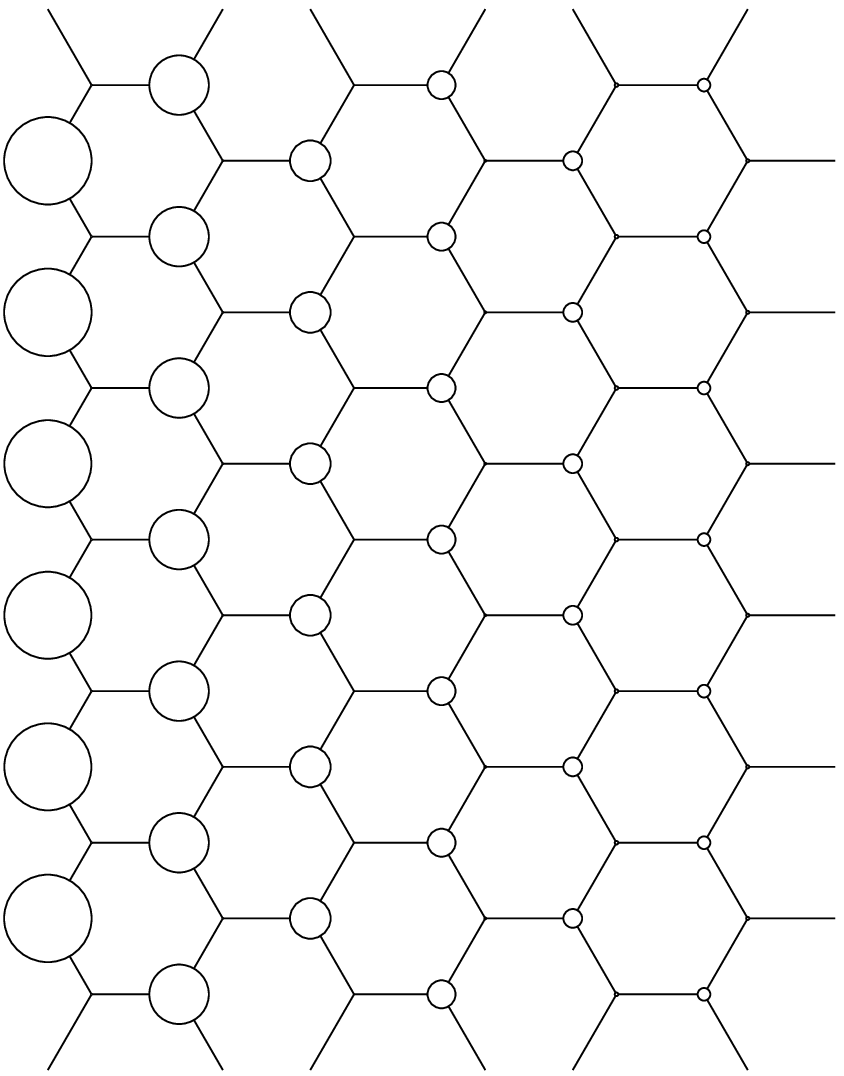}
\hspace{5mm}
\epsfxsize=0.35\hsize
\epsffile{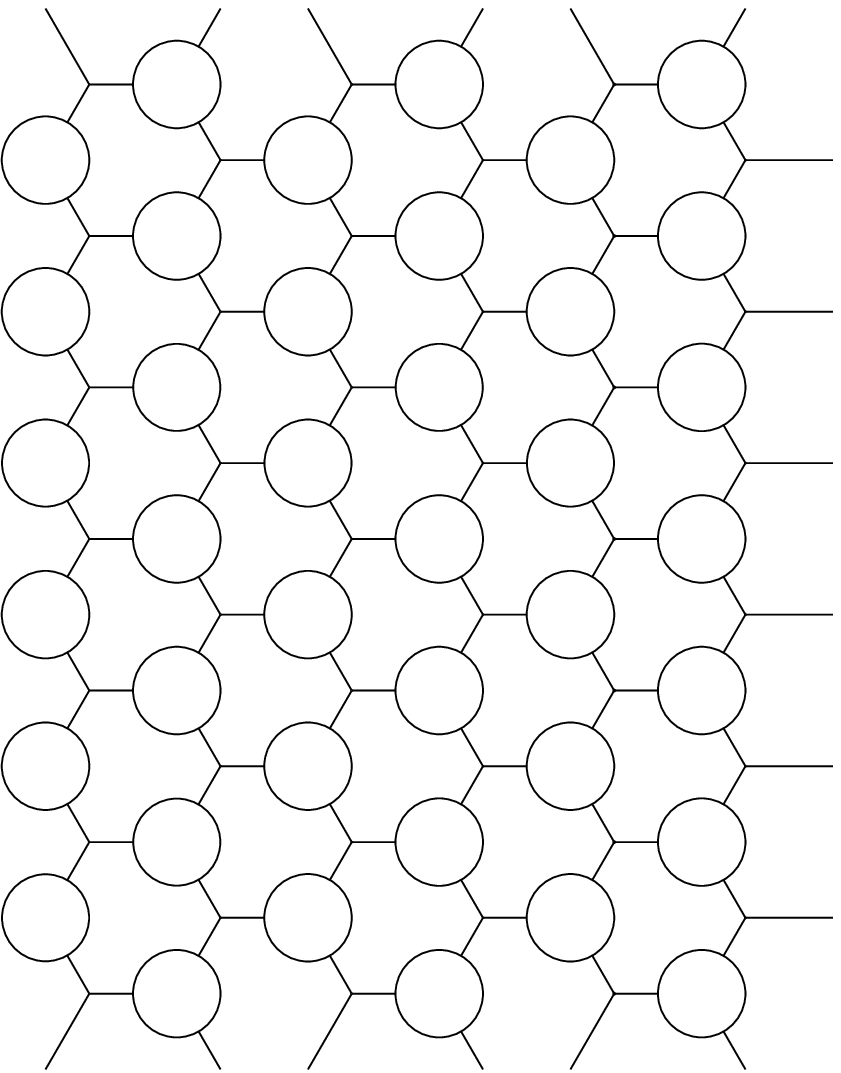}
\caption{The charge density of the edge state at (a) $k=\pi$, (b)
$k=\frac{8}{9}\pi$, (c) $k=\frac{7}{9}\pi$ and (d) $k=\frac{2}{3}\pi$,
where the radius of the circle means the magnitude of the charge
density.} 
\label{fig:charge}
\end{figure}
}

Similarly, the case of finite magnetic field,
using Eqs.(\ref{eq:tmatrix}) and (\ref{eq:init}) under
the condition of $\epsilon=0$, we can derive the wave function  on
the n-$th$ zigzag line as

\begin{eqnarray}
  \Psi_n =\Pi_{i=1}^n D_k(i),
\label{eq:estat_h}
\end{eqnarray}

\noindent
where

\begin{eqnarray}
  D_k(i) = 
\left\{
\begin{array}{cc}
-2\cos\left(\frac{k}{2} - \frac{N-2i+1}{2}\pi\phi\right) &  
(i\geq  2)  \\
1 & (i=1)
\end{array}
\right. 
\end{eqnarray}

\noindent
For the rational flux $\phi = \frac{p}{q}$, 
there is a relation of $D_k(i+q) = D_k(i)$, so that in the case of
semi-infinite graphite Eq.(\ref{eq:estat_h}) might be rewritten as

\begin{eqnarray}
\Psi_{nq} = \left(\Delta_k\right)^{n-1},
\end{eqnarray}

\noindent
where 

\begin{eqnarray}
\Delta_k = \Pi_{i=1}^q D_k(i)
\end{eqnarray}

\noindent
Thus, the edge states are modified in the presence of a magnetic
field. The condition of the convergence of wave function becomes
$\Delta_k\leq 1$, which then defines the region of the flat band.
Note that $\Delta_k$ has q
internal degrees of freedom. In other words, there are q solutions
which give the same value of $\Delta_k$. For example, in the case of
$\phi=\frac{1}{4}$, there are 4 wave numbers which give $\Delta_k=0$,
i.e, $k=\pm\frac{3}{4}\pi,\pm\frac{1}{4}\pi$, the charge density
corresponding to each wave number is depicted in
Fig.~\ref{fig:charge_h}. The charge density
does not penetrate to the inner sites farther than up to the (4+1)-th
zigzag chain, and there are 4 kinds of localized modes corresponding
to the 4 internal degrees of freedom of $\Delta_k$. Thus, 
the edge states in a magnetic field behave as  the
zero-field edge states with q internal degrees of freedom.

{\narrowtext
\begin{figure}[t]
(a) \hspace{30mm}(b)\\
\hspace{5mm}
\epsfxsize=0.35\hsize
\epsffile{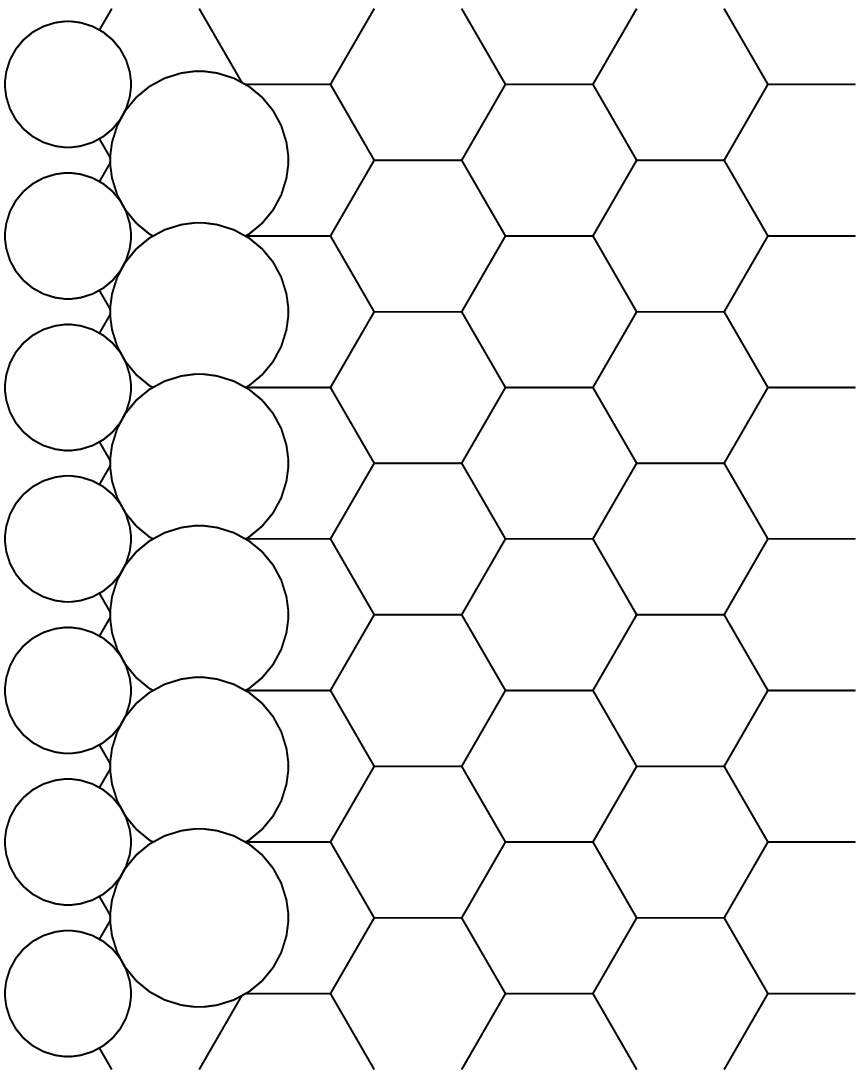}
\hspace{5mm}
\epsfxsize=0.35\hsize
\epsffile{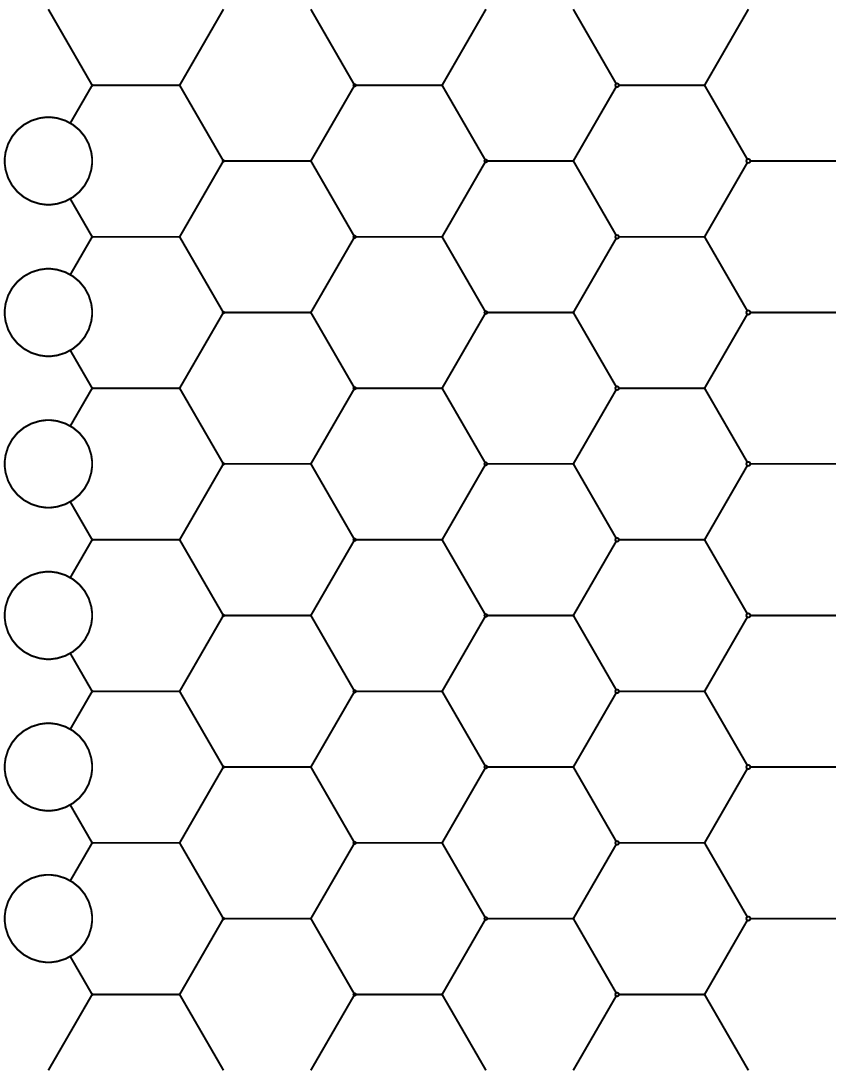}\\
(a) \hspace{30mm}(b)\\
\hspace{5mm}
\epsfxsize=0.35\hsize
\epsffile{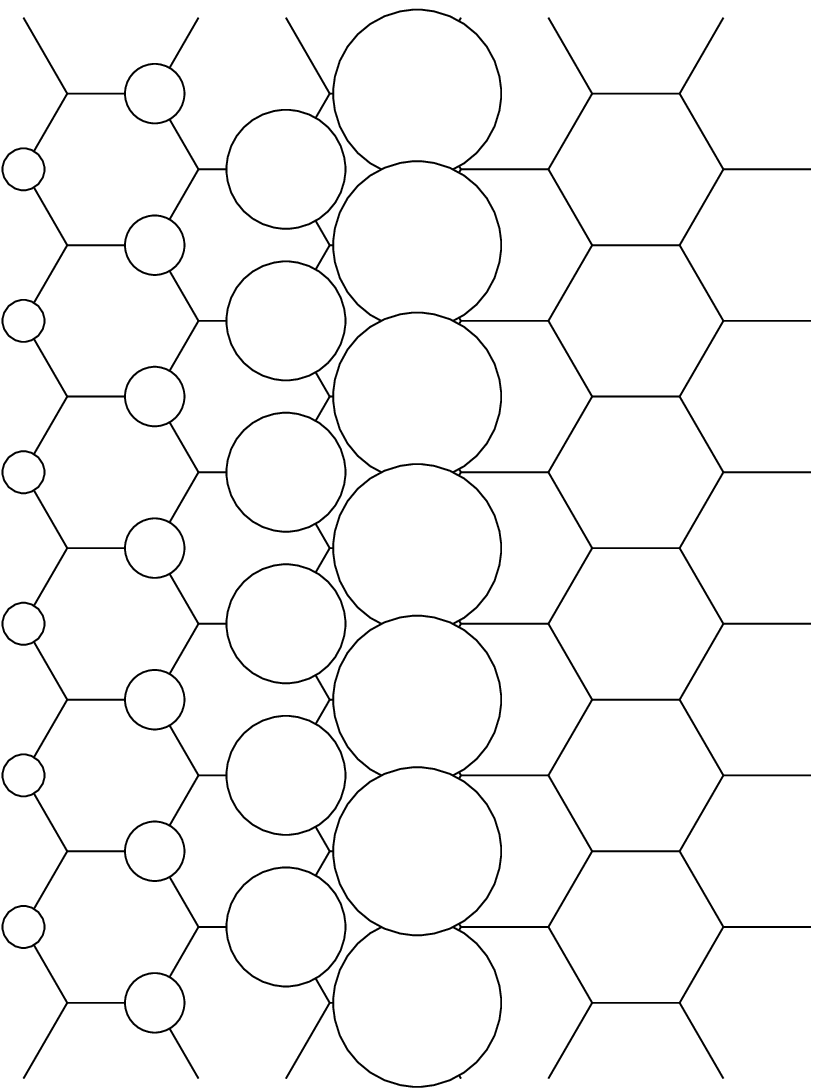}
\hspace{5mm}
\epsfxsize=0.35\hsize
\epsffile{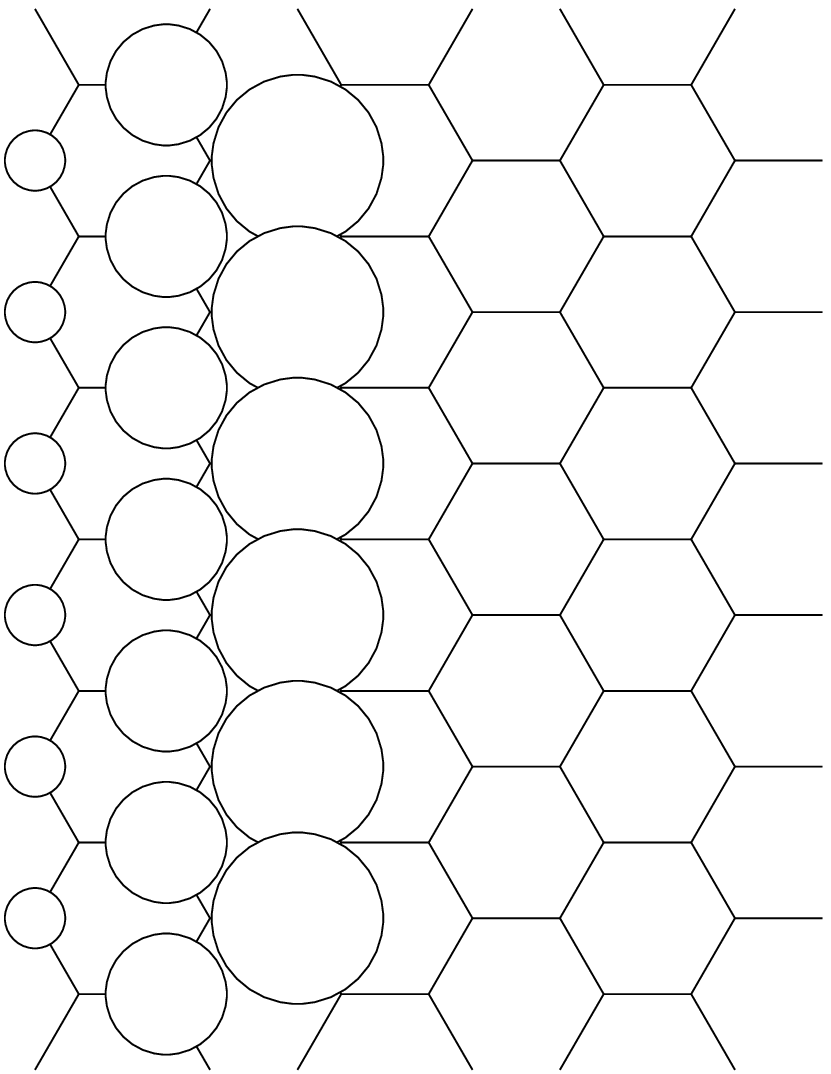}
\caption{The charge density of the edge state at (a)
$k=-\frac{3}{4}\pi$, (b) $k=-\frac{1}{4}\pi$, (c) $k=\frac{1}{4}\pi$
and (d) $k=\frac{3}{4}\pi$, where the radius of the circle denotes the
magnitude of the charge density.} 
\label{fig:charge_h}
\end{figure}
}

Next, we discuss the DOS of the edge state in the absence of a
magnetic field, which will be used in the calculation of the Pauli
susceptibility of zigzag ribbons in the later section. As we have seen
in this section, the edge state penetrate to inner sites when the wave
number changes from $\pi$ to $\frac{2}{3}\pi$. If we consider the
graphite ribbons with width $N$,  two edge states which come from both
side of edge will overlap with each other and develop the bonding and
anti-bonding configurations. Since the magnitude of the overlap becomes
larger when the wave number approaches $\frac{2}{3}\pi$, the band gap
between the bonding and anti-bonding state formed by the two edge
states gets larger toward $k=\frac{2}{3}\pi$. Therefore the partly
flat bands have a slight  dispersion which depends on the ribbon width
$N$. In order to calculate the DOS, we first have to derive the
precise energy dispersion for the edge states. The energy dispersion
is calculated by the overlapping of two edge states. From
Eq.(\ref{eq:estate}), the amplitude of the edge state which penetrates
from the first zigzag line is given by 

\begin{eqnarray}
  \Psi_n =D_k^{n-1}\equiv\Psi_A,
\label{eq:conv}
\end{eqnarray}

\noindent
which is located only on the A-sublattice. On the other hand,the
amplitude of the edge state which penetrate from $N-th$ zigzag line,
is given by

\begin{eqnarray}
  \Psi_{N-n} =D_k^{n-1}\equiv\Psi_B,
\end{eqnarray}

\noindent
which is located only on the B-sublattice. 
By using the tight binding Hamiltonian, the overlapping of two edge
states is easily  calculated,

\begin{eqnarray}
\langle \Psi_A | H | \Psi_B \rangle = N D_k^{N-1}\left( -2t -t
D_k\right)  = T_k, 
\end{eqnarray}

\noindent
where $D_k = -2\cos(\frac{k}{2})$. 
Therefore, the energy spectrum of the edge states is given by the
following eigenvalue problem.

\begin{eqnarray}
 \left(
  \begin{array}{cc}
    0 & T_k \\
    T_k & 0 
  \end{array}
 \right)
 \left(
  \begin{array}{c}
    C_1 \\
    C_2
  \end{array}
 \right)
= \epsilon_k
 \left(
  \begin{array}{c}
    C_1 \\
    C_2
  \end{array}
 \right)
\end{eqnarray}

\noindent
By diagonalization of this Hamiltonian matrix, we find the energy spectrum

\begin{eqnarray}
  E_k = -2tND_k^{N-1}\left( -2t + 2t\cos\left(\frac{k}{2}\right)\right).
\end{eqnarray}

\noindent
From this equation, around $k=\pi$, the spectrum is given by $E\sim k^N$. 

Therefore, the DOS related to the edge states has the form, 

\begin{eqnarray}
\rho(\epsilon)=\frac{\partial k}{\partial \epsilon} \sim
\frac{1}{N}\epsilon^\alpha ,
\label{eq:dos}
\end{eqnarray}

\noindent
where  $\alpha = \frac{1}{N}-1$. Note that this DOS has 
a power-law
dependence, which is different from the ordinary van Hove singularity
of $\rho \sim \frac{1}{\sqrt{E}}$ observed in one-dimensional system. 
It is also found that the renormalized DOS is inversely proportional
to the ribbon 
width, which has been already confirmed by numerical
calculation.\cite{nakada}

\section{orbital diamagnetism of graphite ribbons}
It is well known that graphite shows a large anisotropic diamagnetic
susceptibility, while aromatic molecules show only weak
diamagnetism. 
This fact tells us that the orbital diamagnetic susceptibility is
sensitive to the size of graphite fragments.
Therefore, we would like to clarify the size and edge shape
effect on the orbital susceptibility, in order to understand the
magnetic properties of nanographite ribbons.
Similar calculations of the orbital diamagnetic susceptibility
on carbon nanotubes have been done by several authors\cite{ajiki,Lu}.
They found that there are universal scaling rules in the orbital 
susceptibility as function of the Fermi energy, the temperature
and the size of the nanotubes. 
Since it is also expected that graphite ribbons have such scaling
rules, we study the scaling properties of the orbital susceptibility
and the dependence on the edge shape.

In this section, the orbital diamagnetic susceptibility $\chi_{orb}$
of graphite ribbons is calculated in terms of the 2nd derivative of
the free energy $F(H,T)$ with respect to the magnetic field. 
The free energy is  given by

\begin{eqnarray}
  F(H,T) & = & \mu N -\frac{1}{\beta\pi}\int_{BZ}{\rm d}k\sum_n 
ln\left( 1 + {\rm e}^{-\beta\left(\epsilon_{k,n}(H) - \mu\right) }
\right) \nonumber\\
& &
\end{eqnarray}

\noindent
where $\beta = \frac{1}{k_BT}$ and $\mu$ is the chemical potential and
$\epsilon_{k,n}(H)$ (n is the band index) is the energy spectrum
of the graphite ribbons in the magnetic field, as calculated 
in the previous section.
Then the magnetic moment $M(H)$ and magnetic susceptibility $\chi(H)$
per site
for finite temperature and arbitrary magnetic field $H$ become

\begin{eqnarray}
  M(H)& =& -\frac{1}{N_e}\frac{\partial F}{\partial H} =
-\frac{1}{N_e\pi} \int {\rm d}k \sum_n 
 \frac{1}{{\rm e}^{\beta (\epsilon_{k,n}-\mu)} + 1}
 \frac{\partial \epsilon_{k,n}}{\partial H}, \nonumber\\
& &
\end{eqnarray}

\noindent
and

\begin{eqnarray}
  \chi(H) = \frac{1}{N_e}\frac{\partial M}{\partial H} & = &
-\frac{1}{N_e\pi} \int {\rm d}k \sum_n \left\{
-\frac{\beta}{4\cosh^2\frac{\beta x}{2}}
 \left(\frac{\partial \epsilon_{k,n}}{\partial H}\right)^2
\right.\nonumber\\
&&+
\left.
\frac{1}{{\rm e}^{\beta (\epsilon_{k,n}-\mu)} + 1}
 \frac{\partial^2 \epsilon_{k,n}}{\partial H^2}
 \right\}. \nonumber \\
 & &
\end{eqnarray}

\noindent
The zero-field magnetic susceptibility $\chi_0(T)$ for finite
temperature is given by

\begin{eqnarray}
\chi(T)& = &-\frac{1}{N_e\pi}\left(\frac{S}{\phi_0}\right)^2
 \int {\rm d}k \sum_n^{occ.}
 \frac{1}{{\rm e}^{\beta (\epsilon_{k,n}-\mu)} + 1}
 \left(\frac{\partial^2 \epsilon_{k,n}}{\partial H^2}\right),
\nonumber\\
&&
\end{eqnarray}

\noindent
where $S$ is the area of a hexagonal ring.
$N_e$ is the electron number in the system.

The width dependence of the orbital susceptibility $\chi_{orb}$
at $T=0$ is 
shown in Fig.~\ref{fig:chi_w_0}. Analogous to the graphite sheets, the 
aromatic molecules or the carbon nanotubes, the graphite ribbons
exhibit diamagnetism. The magnitude of $\chi_{orb} (T \approx 0 ) $ grows
linearly with increasing $ W $ in accordance with the fact that $
\chi_{orb} $ of the graphite sheet diverges in the zero-temperature
limit. A remarkable point is the different slope 
in the $W$-dependence of $\chi_{orb}$ for 
armchair and zigzag ribbons. Actually the difference between the
susceptibilities of the two types of ribbons increases for larger $ W
$. At a first glance, this result may seem unphysical, because one
might attribute the difference to an edge effect which should diminish
for wider ribbons. The origin of this discrepancy, however,
is based on topological properties as we will show shortly. We would
like also to mention 
the aspect that $\chi_{orb}$ for the armchair ribbons shows some
oscillations as a function of  $W$. This is due to the fact that
armchair ribbons are metallic or insulating depending on $ W$. On the
other hand, no oscillations occur for zigzag ribbons, as they are
metallic for all $W$.

{\narrowtext
\begin{figure}
\epsfxsize=0.8\hsize
\epsffile{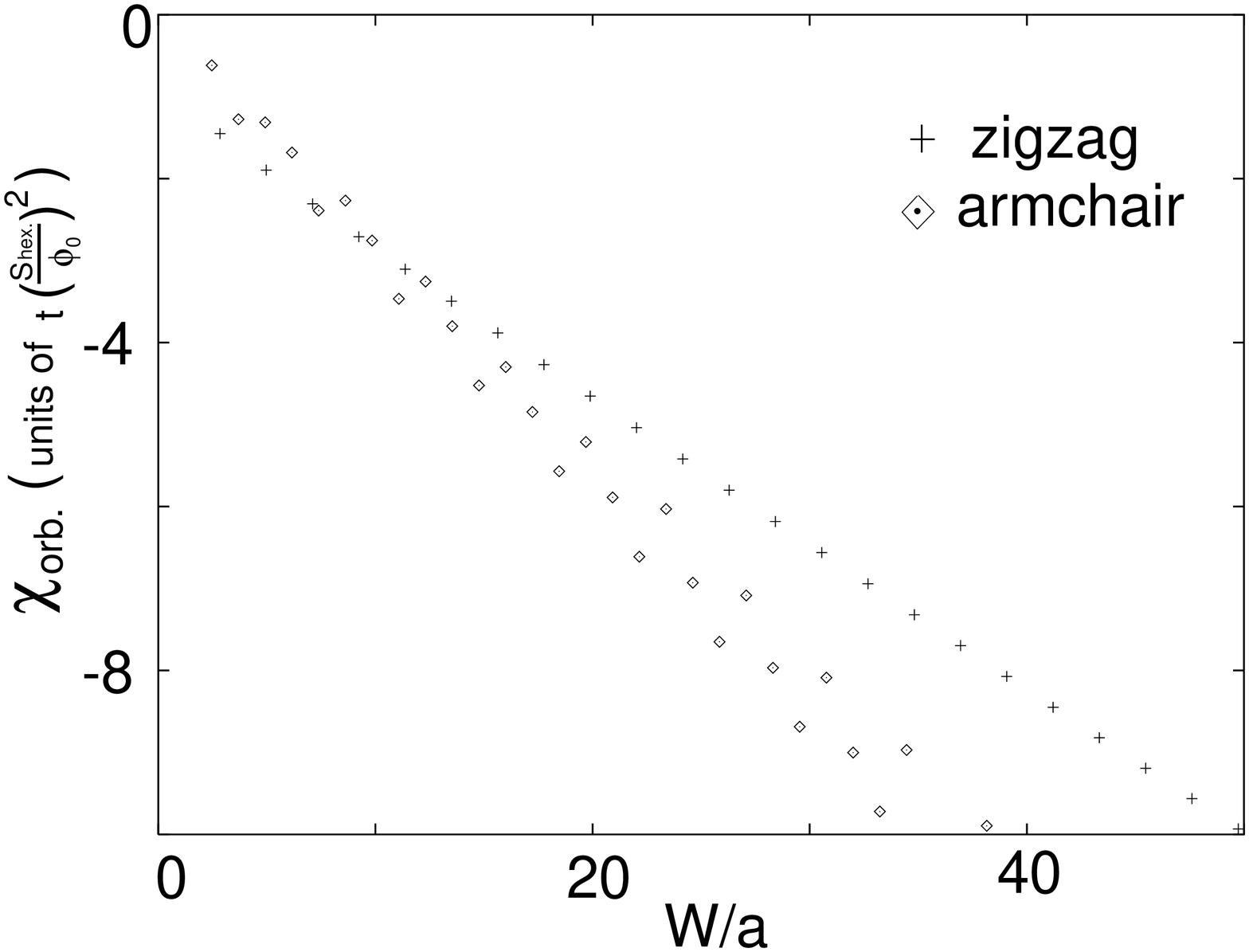}
\caption{The ribbon width dependence of the orbital diamagnetic
susceptibility $\chi_{orb}$ of graphite ribbons at T=0.}
\label{fig:chi_w_0}
\end{figure}
}

The difference of the slope in $\chi_{orb}$ and $M_{orb}$ is related 
to the more microscopic and configurational aspect of the ribbons.
An important point to take into account is the ring current
susceptibility in the equilibrium state, because the magnetic 
moments and ring currents are related in the following way

\begin{equation}
M = \frac{1}{c}\int {\rm d}\mbf(V)\thinspace \mbf(r) \times \mbf(j),
\end{equation}

\noindent
where $\mbf(r)$ is the position and $\mbf(j)$ is the
current operator. $\int{\rm d}\mbf(V)$ means volume integral.
 The current $\mbf(j)$ is described on each bond $(i,j)$ 
by 

\begin{equation}
J_{ij} = {\rm i}\frac{et}{\hbar}\left({\rm e}^{{\rm
i}2\pi\phi_{ij}} c^\dagger_ic_j- {\rm h.c.} \right)
\end{equation}

\noindent
The ring currents also contribute to the linear response for weak
magnetic fields. It is straightforward to calculated them and the
corresponding susceptibility.
The pattern of the ring currents for (a)
a zigzag ribbon of N=10, and armchair ribbons of (b) N=18, (c) N=19
and (d) N=20 are shown in
Fig.~\ref{fig:rc_texture}(a)-(d), respectively. 
The magnitude (in the units 
of $J_0(=\frac{et}{\hbar})$) and directions of the ring
current near the graphite 
edge is further given in detail in Fig~\ref{fig:rc_magnitude_edge}.
We can easily find that the current
flow is symmetric  with respect to $x=0$ and the total current in
$y$-direction 
vanishes, because we consider an equilibrium state.
The pattern of the current flow is strikingly different for the zigzag
and the armchair ribbons.
In zigzag ribbons, due to the lattice symmetry,
the currents along the vertical bonds are exactly zero. 
The currents flow only along the horizontal bonds, whose directions
are antisymmetric with respect to $x=0$.
For armchair ribbons, the current distribution is quite different,
because currents flow also on vertically oriented bonds,   
and exhibits a clear Kekul\'{e}-type of pattern. 
Note that the Kekul\'{e} pattern is more pronounced when
N$\neq$3M-1(M=1,2,3,$\cdots$), which corresponds to the semiconducting armchair
ribbon, whereas it is less distinct for the metallic case, $N=
3M-1$(M=1,2,3,$\cdots$), where  almost no currents are found close to
the ribbon center. The reason can be attributed to the interference
effect between the  current flows assosiated with the two edges. 
As shown in Fig.~\ref{fig:rc_magnitude_edge}(b), the
currents are stronger along the cis-polyacetylene which is
separated by one dimer lines.
For $N\neq 3M-1$ and  $N = 3M-1$, the ring current patterns are depicted
schematically in Fig.~\ref{fig:rc_inter} (a) and (b), where 
thick bold lines denote dominant right-going currents while 
thick shaded lines are for the left-going currents.
It is easy to find that for $N\neq 3M-1$ both types of lines tend to
avoid each other and form a Kekul\'{e} pattern around the center of
the ribbons.
However, when  $N=3M-1$, the lines lie perfectly on top of each other
so that the left- and
right-going currents cancel each other around the center of ribbon.
Thus the effect of the lattice topology near graphite edge
drastically changes the ring current flow in the whole sample.

{\narrowtext
\begin{figure}
(a) \hspace{30mm} (b)\\
\hspace{5mm}
\epsfxsize=0.35\hsize
\epsffile{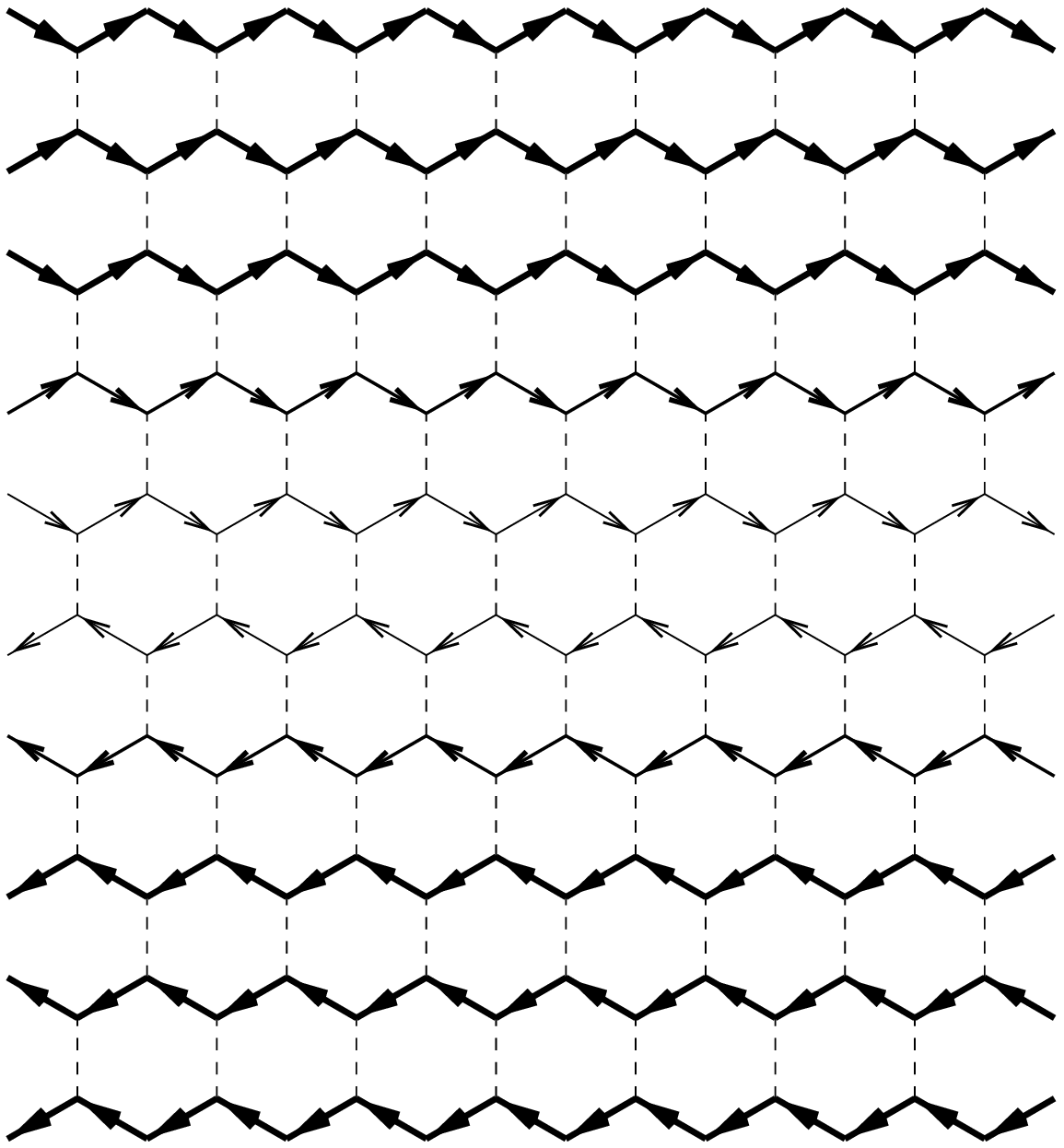}
\hspace{5mm}
\epsfxsize=0.35\hsize
\epsffile{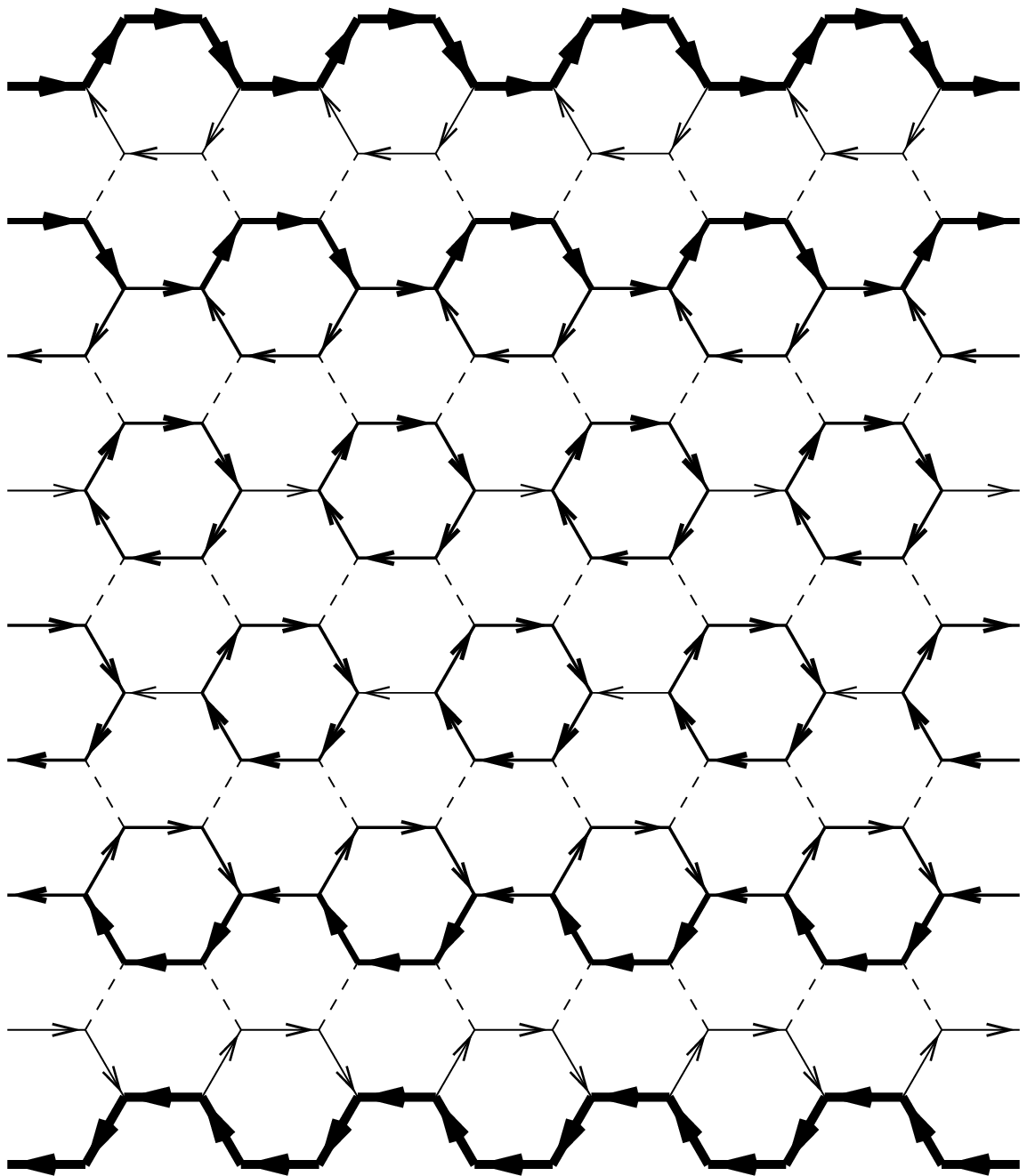}\\
(c) \hspace{30mm} (d)\\
\hspace{5mm}
\epsfxsize=0.35\hsize
\epsffile{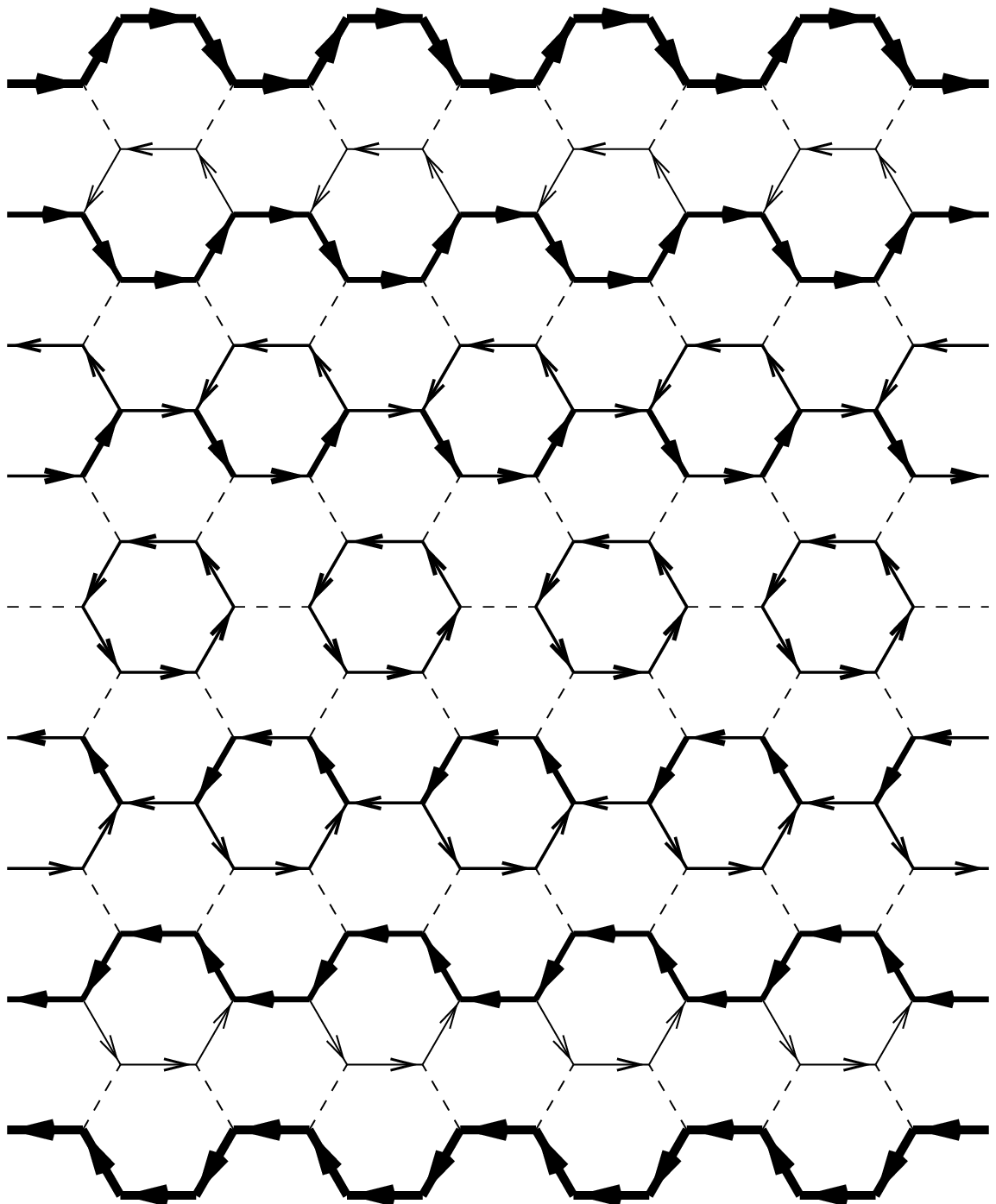}
\hspace{5mm}
\epsfxsize=0.35\hsize
\epsffile{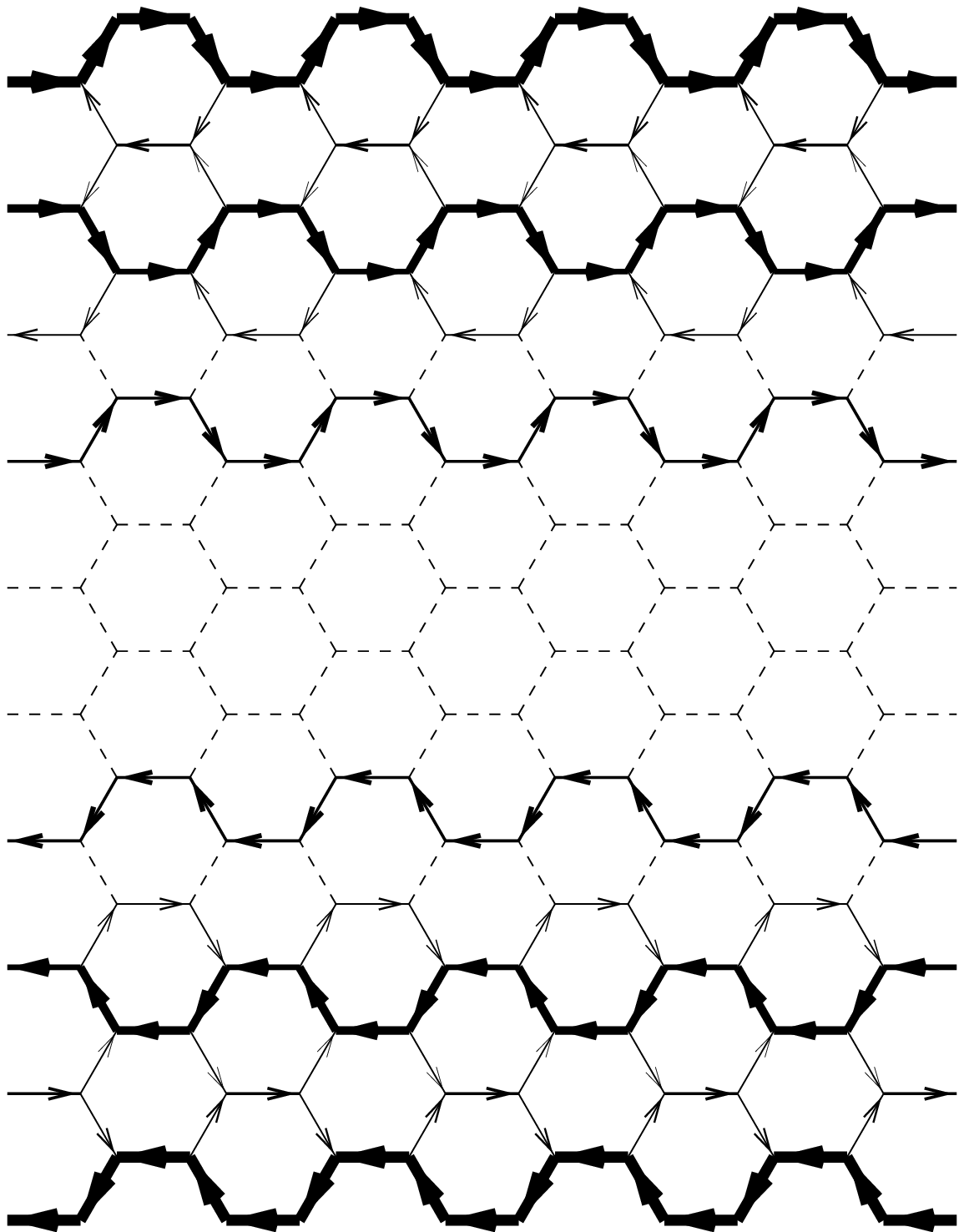}
\caption{The texture of the ring currents for (a) zigzag ribbon
(N=10) and armchair ribbons of (b) N=18, (c) N=19 and (d) N=20.
In zigzag ribbon, because of the symmetry of the lattice, the ring
currents along the vertical bonds are zero. In armchair ribbons of 
N=18 and 19, the Kekul\'{e} pattern is clear. }
\label{fig:rc_texture}
\end{figure}
}

{\narrowtext
\begin{figure}
(a) \hspace{30mm}(b)\\
\hspace{5mm}
\epsfxsize=0.35\hsize
\epsffile{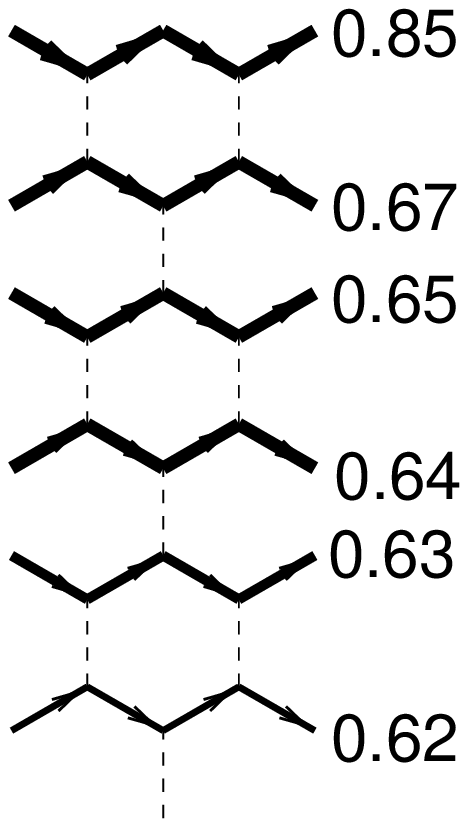}
\hspace{5mm}
\epsfxsize=0.35\hsize
\epsffile{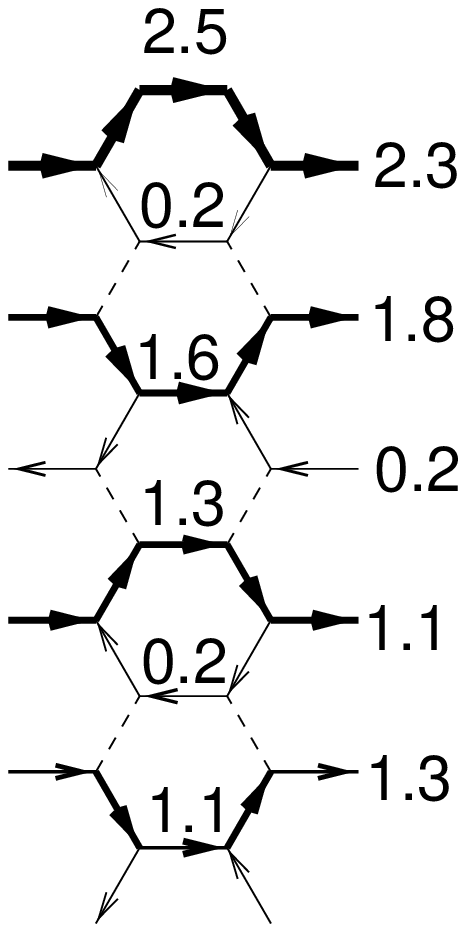}
\caption{Current flow and magnitude near the edge for  
(a) zigzag ribbon (N=50)  and armchair ribbon (N=50). }
\label{fig:rc_magnitude_edge}
\end{figure}
}

{\narrowtext
\begin{figure}
(a) \hspace{30mm}(b)\\
\epsfxsize=0.35\hsize
\epsffile{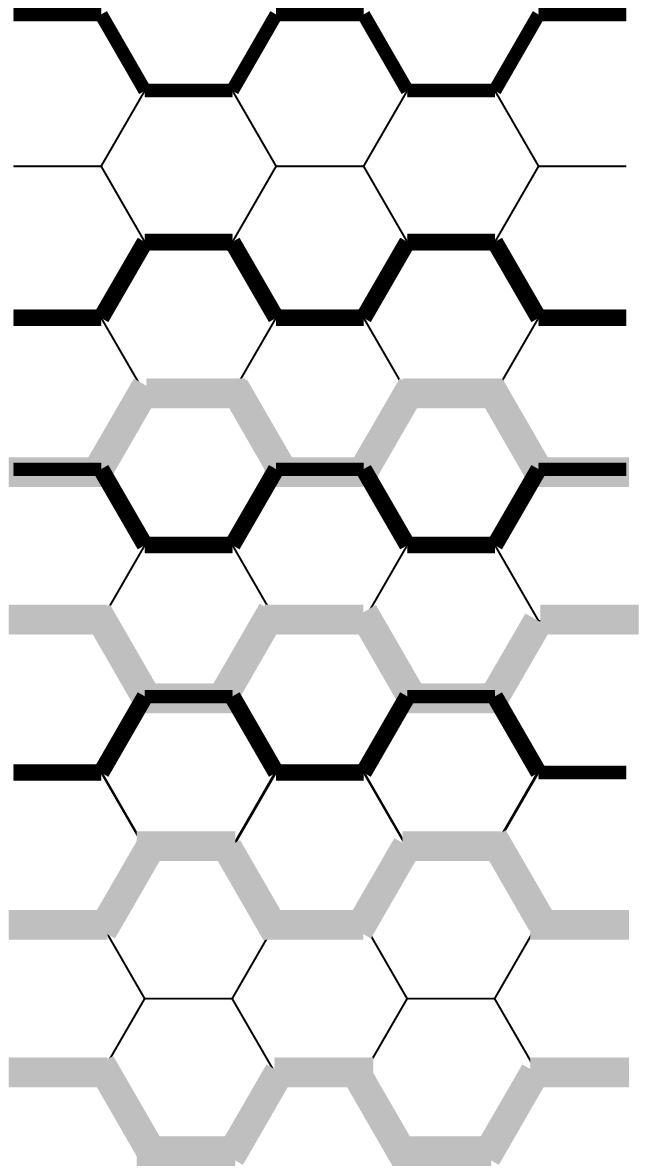}
\hspace{5mm}
\epsfxsize=0.35\hsize
\epsffile{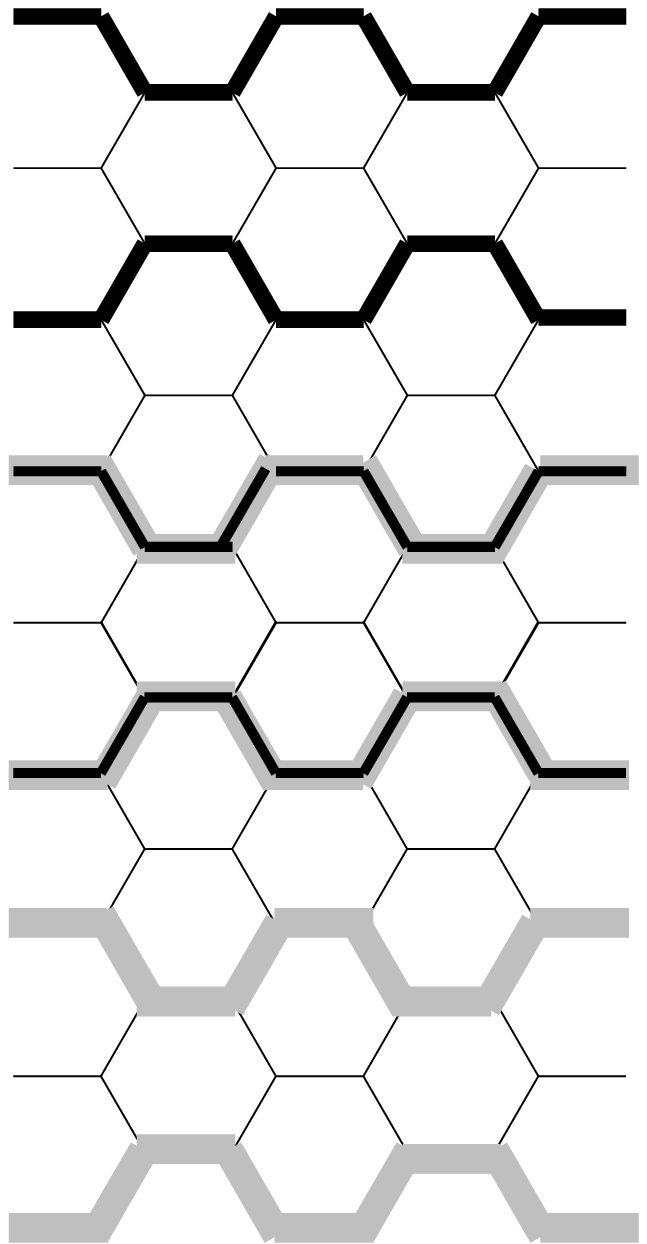}
\caption{Schematic picture of ring current flow generated by the
interference.}
\label{fig:rc_inter}
\end{figure}
}

{\narrowtext
\begin{figure}
(a)\hspace{25mm}(b)\hspace{25mm}(c)\\
\epsfxsize=0.3\hsize
\epsffile{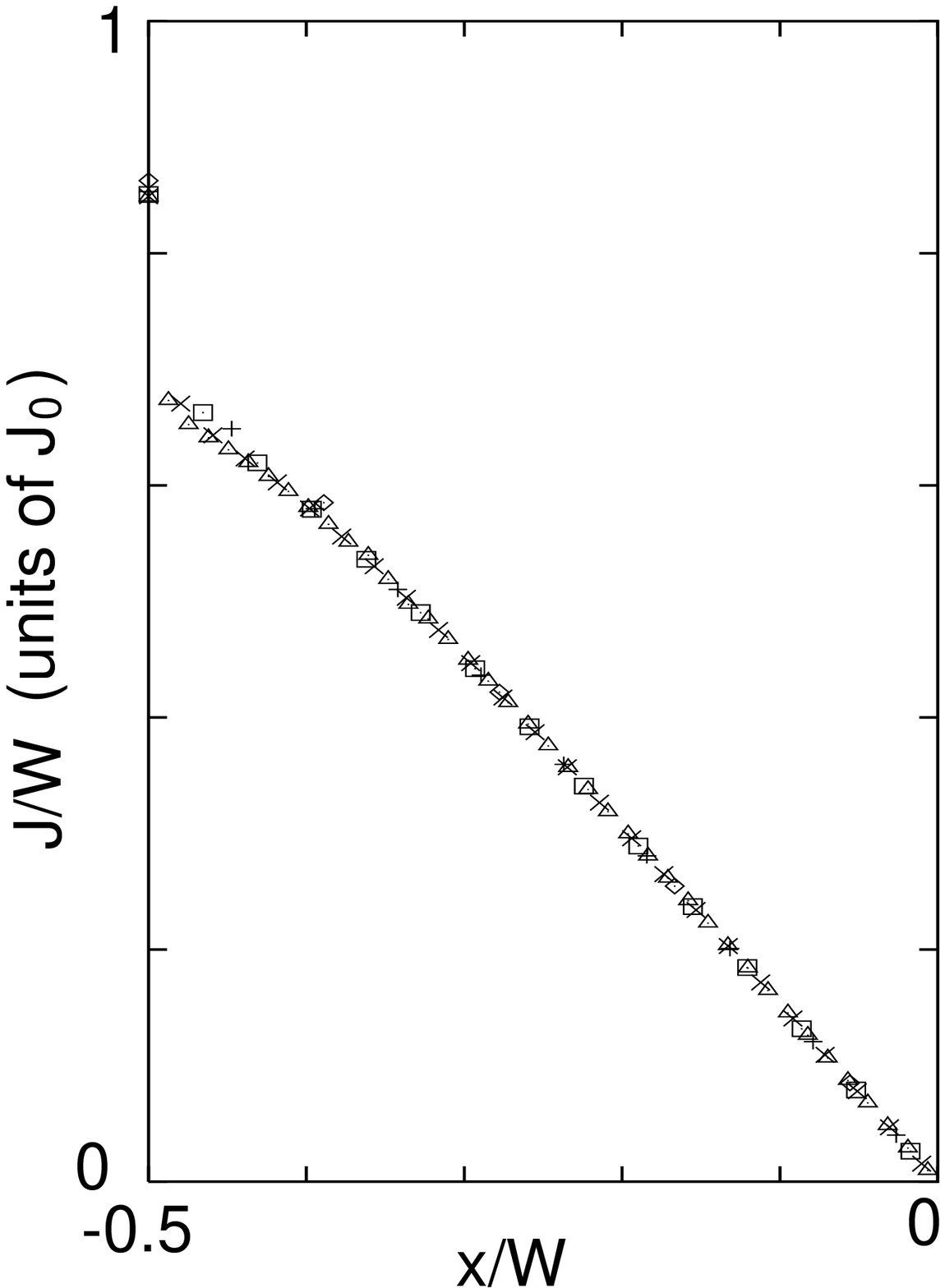}
\epsfxsize=0.3\hsize
\epsffile{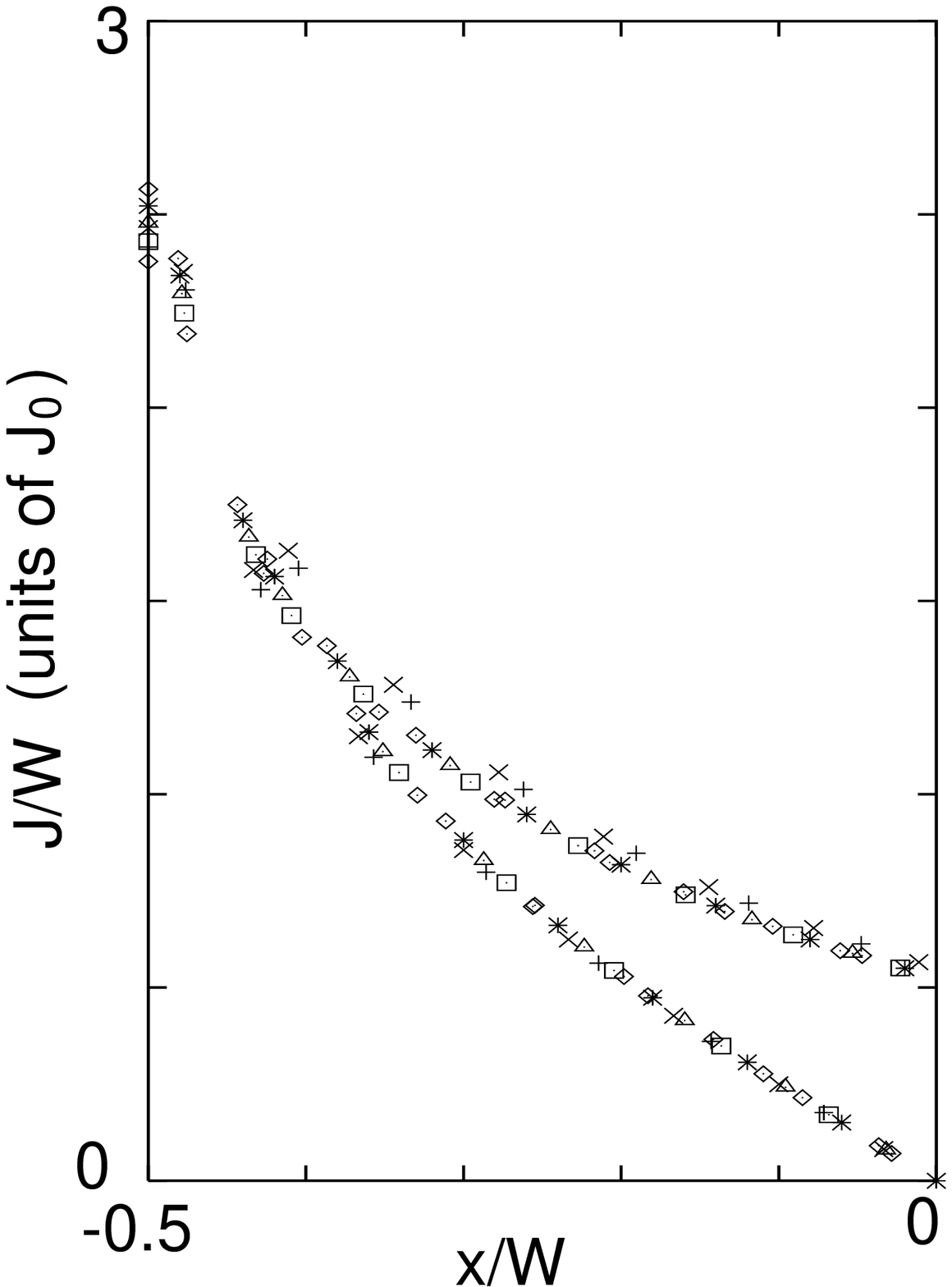}
\epsfxsize=0.3\hsize
\epsffile{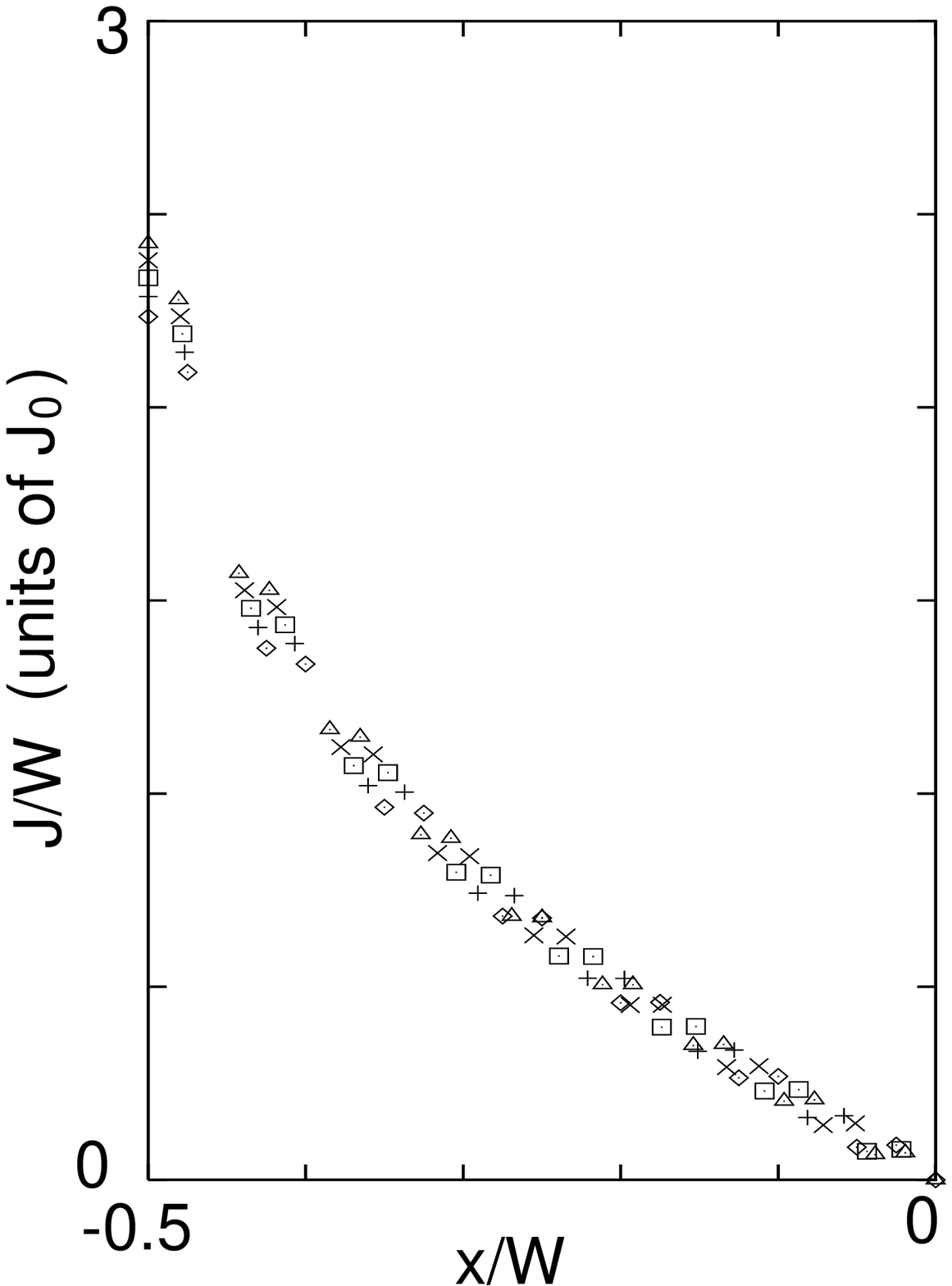}
\caption{The position dependence of magnitude of the ring currents for 
(a) zigzag ribbons, (b) armchair ribbons ($N\neq 3M-1$) and
(c) armchair ribbons ($N= 3M-1$). }
\label{fig:rc_magnitude}
\end{figure}
}

In Fig.\ref{fig:rc_magnitude}, the magnitude of the ring current
susceptibility $J$ is plotted as a function of $x/W$ with a fixed
position $ y $ 
for (a) zigzag ribbons, (b) armchair ribbons with ($N\neq 3M-1$) and
(c) armchair ribbons with ($N= 3M-1$). 
Interestingly, each graphite ribbon has a scaling behavior
as a function of $x/W$ and the magnitude of the ring currents has
a power-law decay.
These facts further emphasize that the edge shape effect is significant
in nanographites.

Next we show the Fermi energy dependence of $\chi_{orb}$.
Actually in real graphite materials, a small change in the  carrier
density from the half-filling is possible and can even be controlled
by substrate properties.
The calculated Fermi energy dependence is shown in
Fig.~\ref{fig:chi_mu_0}, where it is found that $\chi_{orb}/W$ is a
universal function of $\mu W$. We normalize $\chi_{orb}$ by dividing
it by $W$, since it proportional to $W$ ( Fig.~\ref{fig:chi_w_0}).
Furthermore we multiply $E_F$ by $W$, because the direct gap at $k=0$
is proportional to the $1/W$ at $k=0$ for armchair ribbons and at
$k=2\pi/3$  for zigzag ribbons as is demonstrated in the Appendix.

{\narrowtext
\begin{figure}
\epsfxsize=0.8\hsize
\epsffile{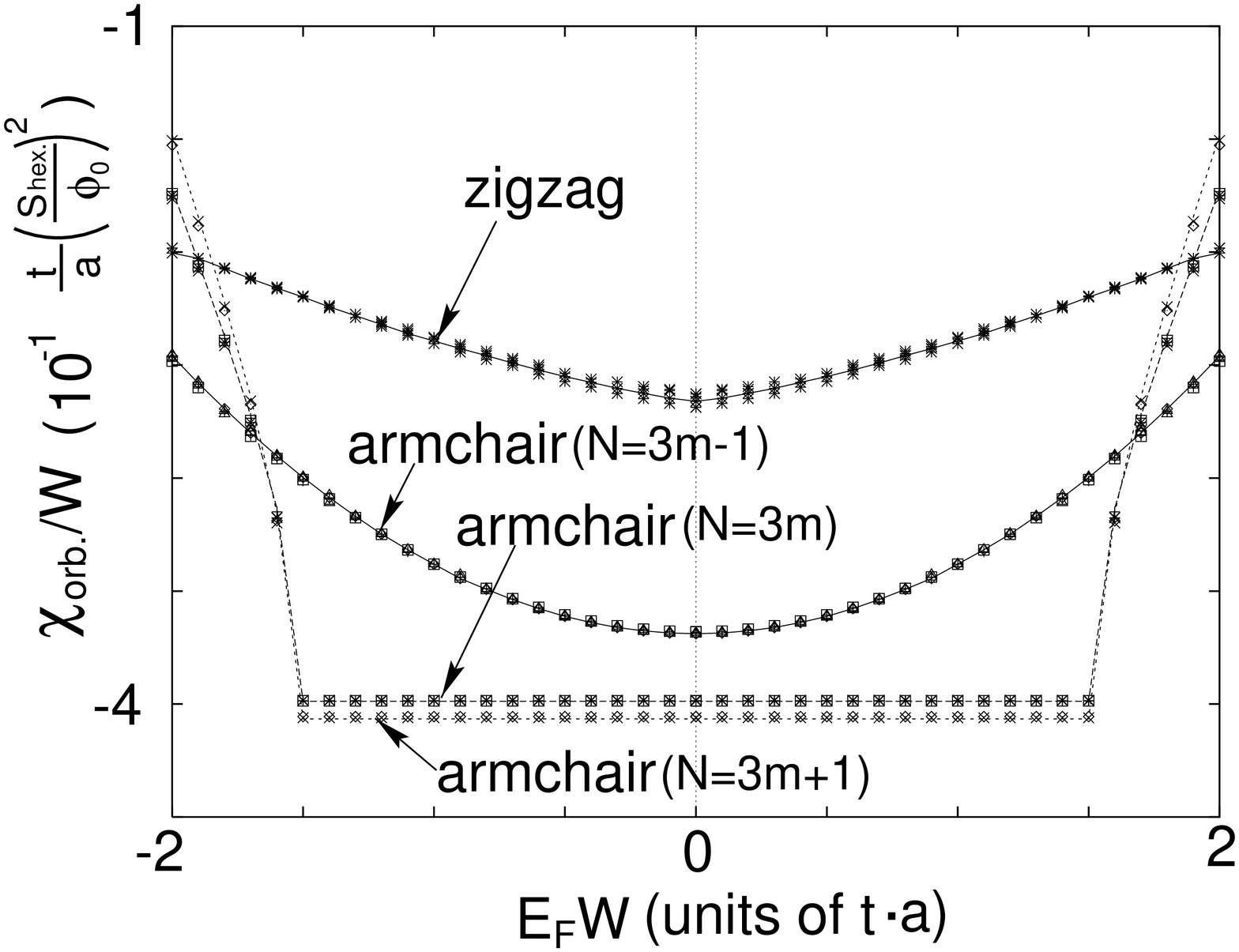}
\caption{The Fermi energy dependence of the orbital magnetic moments 
$\chi_{orb}$ of graphite ribbons at T=0.}
\label{fig:chi_mu_0}
\end{figure}
}

As a final point in this section, we show the temperature dependence 
of $\chi_{orb}$ in Fig.~\ref{fig:orb_temp}, which is important
from the viewpoint of experiments on nanographites.
In all cases the magnitude of $\chi_{orb}$ decreases with increasing
temperature. 
It is also found that the temperature dependence of $\chi_{orb}/W$ 
scales as a function of $\beta W$, because the energy gap is
proportional to the $1/W$. 
Our calculation also demonstrates that the edge effect
becomes more significant with lower temperature.

{\narrowtext
\begin{figure}
\epsfxsize=0.8\hsize
\epsffile{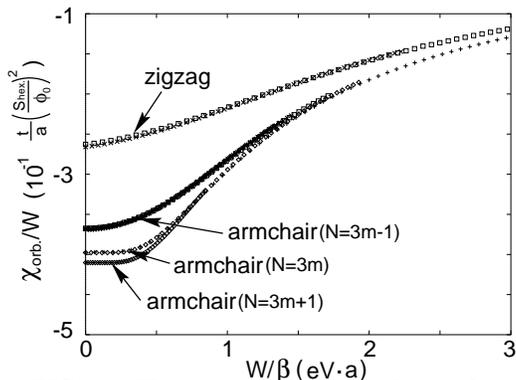}
\caption{The temperature dependence of $\chi_{orb}$, where
$\chi_{orb}$ is scaled by $1/W$ and $\beta$ is scaled by $W$.}
\label{fig:orb_temp}
\end{figure}
}

\section{Pauli Paramagnetism of Graphite Ribbons}
In previous section, we have seen that the orbital diamagnetic 
susceptibility depends on the edge shape in nanographite ribbons, 
especially, the topology of the lattice strongly affects the flow of
diamagnetic ring currents.
Here we discuss another important component of the magnetic
susceptibility, Pauli paramagnetic susceptibility $\chi_P$, 
because zigzag ribbons have a sharp peak of DOS at the Fermi level.
The width of the peak of DOS at the Fermi level has the order of
meV, which is comparable to the temperature scale of room temperature.
Therefore, it is expected that the Pauli susceptibility of zigzag
ribbons might be sensitive to temperature, although the 
Pauli susceptibility of other usual metals is temperature independent.
On the other hand, since the DOS of armchair ribbons at $\epsilon =0$
is zero or very tiny, we can neglect the effect of the Pauli 
paramagnetism  in armchair ribbons.

The magnetic moment by Zeeman effect is

\begin{eqnarray}
  M = \mu_B \left(n_{\uparrow} - n_{\downarrow}\right),
\end{eqnarray}

\noindent
where $\mu_B$ is Bohr magneton and 
$n_{\uparrow}$ ($n_{\downarrow}$) means the electron density with
up-spin (down-spin). 
The electron density at arbitrary temperature for each spin is given by

\begin{eqnarray}
  n_{\sigma} = \frac{1}{\pi}\int_{1st \bf BZ}{\rm d}k
\sum_n \frac{1}{1+{\rm e}^{\beta\left(\epsilon_{n,k}-\sigma \mu_BH\right)}},
\end{eqnarray}

\noindent
where $\sigma(=\uparrow,\downarrow)$ means spin index. 
Therefore, the Pauli susceptibility $\chi_P$ per site is given by

\begin{eqnarray}
  \chi_P = \lim_{T\rightarrow 0} \frac{\partial M}{\partial H} 
         = \frac{\beta\mu_B^2}{\pi N_e}\sum_{n}\int{\rm
         d}k\frac{1}{\cosh\left(\beta\epsilon_{n,k}\right)}, 
\label{eq:pauli}
\end{eqnarray}

\noindent
where $\beta=\frac{1}{k_B T}$. Room temperature ($T\sim 300$K)
corresponds to $\beta \sim 0.25$.
We numerically calculated the finite temperature Pauli susceptibility
of graphite ribbons using this equation up to room temperature.

{\narrowtext
\begin{figure}
\epsfxsize=0.8\hsize
\epsffile{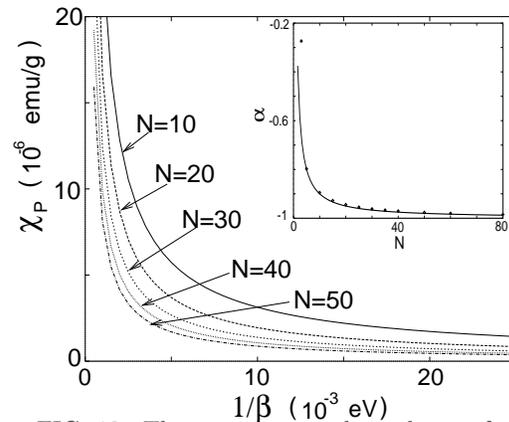}
\caption{The temperature dependence of $\chi_P$ for 
N= 10, 20, .. , 50 up to room temperature.
In the inset, the exponet is plotted on the width N and show a
good agreement with the line of $\frac{1}{N}-1$.}
\label{fig:pauli}
\end{figure}
}

It is possible to clarify the contribution of the edge states to
$\chi_P$. As we have seen in Sec.III, the DOS 
due to the edge states is 
given by Eq.(\ref{eq:dos}).
After the substitution of Eq.(\ref{eq:dos}) into 
Eq.(\ref{eq:pauli}), we replace the k-integration by the energy
integration. Then we can obtain the $\chi_P$ contribution due to the
edge states as,

\begin{eqnarray}
  \chi_P = \frac{1}{N_eN\beta^\alpha}\int{\rm d}x\frac{x^\alpha}
{\cosh x+1} \sim \frac{1}{N}T^\alpha,
\label{eq:pauli_power}
\end{eqnarray}

\noindent
where $x$ is $\beta \epsilon_k$ and $\alpha$ is $\frac{1}{N}-1$.
Interestingly, $\chi_P$ has the Curie-like temperature dependence,
although in normal metals
$\chi_P$ is basically constant in the temperature.
The exponent of $\chi_P$ depends on the ribbon width through 
$\alpha$. 
When $N$ becomes infinite, the exponent $\alpha$ approches $-1$ 
and $\chi_P$ 
show the Curie-law. However, in this limit,  the 
contribution of $\chi_P$ is diminished by a factor $1/N$ 
in Eq.(\ref{eq:pauli_power}).

Numerical results of the Pauli susceptibility $\chi_P$ 
of zigzag ribbons up to room temperature are shown 
in Fig.~\ref{fig:pauli} for various values of $N$. 
As expected, because of the edge states, $\chi_P$ shows 
Curie-like temperature dependence.
In the inset of Fig.\ref{fig:pauli}, we plotted the N dependence of
$\alpha$, which was calculated by the least square method and
has a good agreement with the line of $\frac{1}{N}-1$.

The observed susceptibility $\chi$ is essentially the sum of the 
orbital $\chi_{orb.}$
and the Pauli susceptibility $\chi_P$. 
The temperature dependence of the total susceptibility $\chi$ 
is shown 
in Fig.\ref{fig:pauli+dia}. 
The total susceptibility $\chi$ shows the diamagnetic behavior 
in the high temperature regime and paramagnetic behavior in the low
temperature.
In the inset, the width dependence of the crossing temperature,
i.e. $\chi=0$, is plotted, which is well fitted by
$\frac{1}{\beta}= 10^{0.1526}\times   x^{-1.846}$.

Here we should remind that both aromatic molecules and bulk graphite 
show diamagnetic behavior, howerver, nanographite with zigzag edges
have a remarkable paramagnetic behavior because of the edge state.
If this paramagnetic behavior is experimentally detected,
it will be an indirect evidence of the existence of the edge state.

{\narrowtext
\begin{figure}
\epsfxsize=0.8\hsize
\epsffile{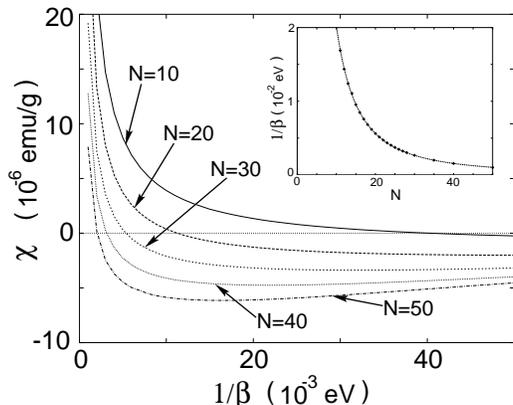}
\caption{The temperature dependence of 
total susceptibility $\chi$, which is $\chi_{orb.} +
\chi_P$ is shown for $N= 10, 20, ... , 50$. In the 
inset, the width dependence of crossing temperature, where
$\chi=0$.}
\label{fig:pauli+dia}
\end{figure}
}

\section{Summary and Discussion}
In this manuscript, we discussed the electronic and magnetic
properties of nanographite in magnetic field, by using  the
single orbital tight binding model with Peierls phase.
Deriving the Harper equation for the graphite lattice,
we studied the energy spectrum and dispersion in a magnetic field.
At $\phi=0$, it is found that zigzag ribbons have
partly flat bands at the Fermi level, which is attributed to
the edge localized states having non-bonding character.
In terms of the Harper equation, the edge state can be analytically 
described even in presence of a magnetic field.
We also studied orbital diamagntic properties of the nanographite
ribbons, where we found that the diamagnetic susceptibility
$\chi_{orb.}$ is very sensitive to the size and edge shapes of
graphite ribbons. It is emphasized that the flow of the orbital
diamagnetic ring currents significantly depend on the lattice 
topology near the graphite edges. 
Especially, in the case of armchair ribbons, the pattern of the ring
currents has drastically changed because of the interference effect of 
two edges. It is also found that the orbital diamagnetic
susceptibility $\chi_{orb.}$ is scaled as a
function of the temperature, Fermi energy and ribbon width.
Because the edge states induce a sharp peak of DOS at the Fermi
level, the Pauli paramagnetic susceptibility should be an
important component in nanographite with zigzag edges.
Therefore, in 
the last section the Pauli susceptibility in zigzag ribbons 
has been studied, where we found that the zigzag ribbons with nanometer
size show the Curie like temperature dependence of the Pauli
susceptiblity in contrast to usual metals.
From this significant contribution of the Pauli susceptibility, 
it is found that the observed susceptibility 
$\chi(=\chi_P+\chi_{orb.})$  of zigzag ribbons show diamagnetic
behavior at high temperature and paramagnetic behavior at low
temperature.

Here we introduce an interesting experimental results, which might
be connected with our theoretical results.
Some graphite-related materials consisting of nanographites, 
{\it e.g.,} activated carbon fibers (ACF), amorphous carbons, 
carbon blacks, defective carbon nanotubes etc., 
show actually anomalous behaviors in the magnetic susceptibility.
While bulk graphite has a large diamagnetic and
anisotropic susceptibility,
a certain type of ACF with 
 a huge specific surface area (SSA) up to 3000$m^2/g$
(believed to consist of an assembly of minute graphite
fragments with a dimension of $20$\AA $\times$ $20$\AA)
exhibits an paramagnetic response at room temperature
and a strong Curie-like behavior in low temperature\cite{nakayama}.
This kind of anomalous behavior of the susceptibility is also observed 
in many amorphous carbons and defective carpet-rolled carbon 
nanotubes\cite{bandow}. 
Although zigzag and armchair edges coexist in real carbon 
material, this behavior of magnetic susceptiblity is consistent with
our results.
Although the sample production of graphite-related materials has still
insufficient influence on size and edge shapes, 
recently there are some experimental attempts to synthesize and
nanographite systems and to control the size and edge shapes. One is
``graphitization'' of diamond powder with grain sizes $40-50$\AA.
Another method to produce
nanographites is epitaxial growth  on substrates with  step
edges\cite{terai}. 
Therefore, we  expect
that in near future the magnetic properties of nanographites 
will be observable so that the influence of the edge states on
magnetic properties can be tested in a controlled way.

\section*{Acknowledgments}
The authors are grateful to K. Kusakabe, H. Tsunetsugu, K. Nakada and
M. Igami for many helpful discussions. 
K.W. acknowledges the Research Fellowships of the Japan Society for
the Promotion of Science for Young Scientists.
This work has been supported 
by a Grant-in-Aid for Scientific Research 
(09875066) 1997
and by a Grant-in-Aid 
for Scientific Research on Priority Areas ``Carbon Alloys'' 
 (09243105) from the Ministry of Education, 
Science and Culture, Japan (M.~F.).

\appendix
\section{Energy Gap}
In this Appendix, we analytically show that 
both the direct gap $\Delta_a$ at $k=0$ of sufficiently wide
armchair ribbons and
the direct gap $\Delta_z$ at $k=\frac{2\pi}{3}$ of sufficiently wide
of zigzag ribbons are inversely proportional to the width of graphite
ribbon $W$. This result supports that the $\chi_{orb}/W$ is scaled as 
a function of the $\mu W$ ($\mu$ is chemical potential).

First, we examine the energy gap $\Delta_a$  at $k=0$ of armchair ribbons.
It is easy to find that at the $k=0$ the Hamiltonian can be rewritten as 

\begin{equation}
H\!=\!-t\sum_{j=1}^{N}\{
\sum_{\mu=1}^{2}(a_{j,\mu}^{\dagger}a_{j+1,\mu} + h.c.)
+a_{j,1}^{\dagger}a_{j,2}+h.c.\},
\end{equation}

\noindent
which is equivalent to the tight binding model for
the 2-leg ladder lattice having N rungs\cite{peculiar}.
The site indices (j,1) and (j,2) correspond to the jA(B) and jB(A) 
sites, respectively, when j is even(odd). The eigenvalues are 
evaluated as $\epsilon^{\pm}=-2t\cos n\pi/(N+1)\pm t$ 
($n$=1,2,\ldots,$N$). It should be noted that
the system is metallic only when $N$=3$m$-1, because $\epsilon^{+}$
and $\epsilon^{-}$ become zero for $n=m$ and $2m$, respectively.
Therefore, $\Delta_a$ are 0 for $N$=3$m$-1, 
$2\left(2t\cos(\frac{m}{3m+1}\pi)-t\right)$ for $N$=3$m$ and
$2\left(2t\cos(\frac{m+1}{3m+1}\pi)-t\right)$ for $N$=3$m$+1,
respectively. After the elimination of N in terms of 
$W=(N-3)\frac{\sqrt{3}}{2}+\sqrt{3}$ and the Taylor expansion under the
condition of $1/W\ll 1$, we can obtain the following results.

\begin{eqnarray}
\Delta_a \sim \left\{ 
\begin{array}{lccc}
0  &             N&=&3M-1\\ 
\displaystyle{\frac{\pi}{W+\frac{\sqrt{3}}{2}}} &  N&=&3M \\ 
\displaystyle{\frac{\pi}{W}} &  N&=&3M+1 
\end{array}
\right.
\end{eqnarray}

\noindent
Thus the $\Delta_a$ is inversely proportional to the ribbon width.

Similarly, we can obtain the energy gap $\Delta_z$  at
$k=\frac{2\pi}{3}$ of zigzag ribbons.
The Hamiltonian of zigzag ribbons at $k=\frac{2\pi}{3}$  is rewritten as 

\begin{equation}
H\!=\!-t\sum_{i=1}^{2N}(a_{i}^{\dagger}a_{i+1} + h.c.),
\end{equation}

\noindent
which is equivalent to the tight binding model for
the one-dimensional lattice having 2N sites.
The site index i corresponds to iA, if i is an odd number,
and to  iB ,if i is an even number.
The eigenvalues are 
evaluated as $\epsilon=-2t\cos n\pi/(2N+1)$ 
($n$=1,2,\ldots,$N$). 
Therefore, $\Delta_z$ is 
$4\left(2t\cos(\frac{N+1}{2N+1}\pi)\right)$.
After the elimination of N in terms of 
$W=\frac{\sqrt{3N}}{2}-1$ and Taylor expansion under the
condition of $1/W\ll 1$, we can obtain the following results.

\begin{equation}
\Delta_z \sim \frac{\pi}{W}
\end{equation}

\noindent
Thus the $\Delta_a$ is also inversely proportional to the ribbon width.

\end{multicols}
\end{document}